\documentclass[runningheads]{llncs}

\usepackage[T1]{fontenc}
\usepackage[utf8]{inputenc}
\usepackage{microtype}
\usepackage{amsmath,amssymb}
\usepackage{booktabs}
\usepackage{algorithm}
\usepackage[noend]{algpseudocode}
\usepackage{longtable}
\usepackage{array}
\usepackage{arydshln}
\usepackage{graphicx}
\usepackage{multirow}
\usepackage{pifont}
\usepackage{newunicodechar}
\newunicodechar{✓}{\ding{51}}
\newunicodechar{✗}{\ding{55}}
\usepackage{fontawesome5}
\newunicodechar{⏱}{\scalebox{0.8}{\faClock[regular]}}
\newunicodechar{❗}{\textbf{!}}
\newunicodechar{🔋}{\scalebox{0.8}{\faBatteryEmpty}}
\usepackage{xcolor}
\usepackage{alltt}
\usepackage{tikz}
\usetikzlibrary{positioning,fit,backgrounds,arrows.meta,calc,shapes.misc}
\usepackage{hyperref}
\usepackage{orcidlink}\renewcommand{\orcidID}[1]{\orcidlink{#1}}
\usepackage[capitalise,nameinlink]{cleveref}

\newcommand{\tool}{\textsc{OmegaHyper}}

\newcommand{\autohyper}{\textsc{AutoHyper}}

\newcommand{\hypersmv}{\textsc{HyperSMV}}
\newcommand{\hyperqb}{\textsc{HyperQB}}
\newcommand{\hyperlasso}{\textsc{HyperLasso}}
\newcommand{\spot}{\textsc{Spot}}
\newcommand{\kofola}{\textsc{Kofola}}
\newcommand{\nusmv}{\textsc{NuSMV}}
\newcommand{\nuxmv}{\textsc{nuXmv}}
\newcommand{\mchyper}{\textsc{MCHyper}}
\newcommand{\divine}{\textsc{DiVinE}}
\newcommand{\rabit}{\textsc{Rabit}}
\newcommand{\bait}{\textsc{Bait}}
\newcommand{\forklift}{\textsc{ForkLift}}
\newcommand{\forq}{\textsc{Forq}}
\newcommand{\roll}{\textsc{Roll}}

\newcommand{\g}[1]{\textcolor{black!40}{#1}}

\newcommand{\flag}[1]{\texttt{-{}-#1}}
\definecolor{exRigidC}{HTML}{C65D00}\definecolor{exInputC}{HTML}{2A6DB5}\definecolor{exDeclC}{HTML}{2E8B3A}
\newcommand{\exRigid}[1]{\textcolor{exRigidC}{#1}}\newcommand{\exInput}[1]{\textcolor{exInputC}{#1}}\newcommand{\exDecl}[1]{\textcolor{exDeclC}{#1}}
\newcommand{\exLine}[1]{\textcolor{black!45}{\makebox[1.3em][r]{\tiny #1}}\hspace{0.5em}}
\newlength{\toolcolw}
\definecolor{fastGroup}{HTML}{FFE699}
\definecolor{fastAll}{HTML}{CFE8C6}
\newcommand{\swatch}[1]{\tikz[baseline=-0.6ex]\node[rounded corners=1pt, minimum width=4mm, minimum height=2.4mm, inner sep=0pt, draw=black!60, fill=#1] {};}
\makeatletter
\newcommand{\cellbg}[1]{%
  \setbox0\hbox{\rowstrut}%
  \dimen0=\ht0 \ifdim\ht\@arstrutbox>\dimen0 \dimen0=\ht\@arstrutbox\fi
  \dimen2=\dp0 \ifdim\dp\@arstrutbox>\dimen2 \dimen2=\dp\@arstrutbox\fi
  \edef\cellbg@x{\the\dimexpr\toolcolw+2\tabcolsep\relax}\edef\cellbg@y{\the\dimexpr\dimen0+\dimen2\relax}%
  \edef\cellbg@dx{\the\tabcolsep}\edef\cellbg@dy{\the\dimen2}%
  \makebox[0pt][l]{\begin{tikzpicture}[overlay,x=\cellbg@x,y=\cellbg@y,shift={(-\cellbg@dx,-\cellbg@dy)}]#1\end{tikzpicture}}}
\makeatother
\newcommand{\fg}[1]{\cellbg{\fill[fastGroup] (0,0) rectangle (1,1);}\makebox[\toolcolw]{#1}}
\newcommand{\fa}[1]{\cellbg{\fill[fastAll] (0,0) rectangle (1,1);}\makebox[\toolcolw]{#1}}
\newcommand{\benchname}[3]{#1\if\relax\detokenize{#2}\relax\else$_{\text{\textsc{#2}}}$\fi\if\relax\detokenize{#3}\relax\else$#3$\fi}
\newcommand{\bakery}[1]{\benchname{Bakery}{}{#1}}
\newcommand{\snark}{\benchname{SNARK}{}{}}
\newcommand{\snarkbug}{\benchname{SNARK}{bug}{}}
\newcommand{\snarkfix}{\benchname{SNARK}{fix}{}}
\newcommand{\nicorrect}{\benchname{NI}{correct}{}}
\newcommand{\niincorrect}{\benchname{NI}{incorrect}{}}
\newcommand{\nrpcorrect}{\benchname{NRP}{correct}{}}
\newcommand{\nrpincorrect}{\benchname{NRP}{incorrect}{}}
\newcommand{\robotrobust}[1][]{\benchname{Robot}{robust}{\if\relax\detokenize{#1}\relax\else#1^2\fi}}
\newcommand{\robotshort}[1][]{\benchname{Robot}{short}{\if\relax\detokenize{#1}\relax\else#1^2\fi}}
\newcommand{\robottwo}{\benchname{Robot}{two}{}}
\newcommand{\robotscore}{\benchname{Robot}{score}{}}
\newcommand{\robotadv}{\benchname{Robot}{adv}{}}
\newcommand{\robotplan}[2]{\benchname{Robot}{plan}{#1\!\times\!#2}}
\newcommand{\mutant}{\benchname{Mutant}{}{}}
\newcommand{\coterm}{\benchname{CoTerm}{}{}}
\newcommand{\deniability}{\benchname{Deniability}{}{}}
\newcommand{\bufscod}{\benchname{BufS}{cod}{}}
\newcommand{\bufsigmni}{\benchname{BufS}{igmni}{}}
\newcommand{\bufsiod}{\benchname{BufS}{iod}{}}
\newcommand{\bufucod}{\benchname{BufU}{cod}{}}
\newcommand{\nodettini}{\benchname{Nodet}{tini}{}}
\newcommand{\nodettsni}{\benchname{Nodet}{tsni}{}}
\newcommand{\ksafety}{\benchname{KSafety}{}{}}
\newcommand{\mapsynth}[1]{\benchname{MapSynth}{}{#1}}
\newcommand{\teamltl}[1]{\benchname{TeamLTL}{}{#1}}
\newcommand{\ndet}[1]{\benchname{NDET}{}{#1}}
\newcommand{\bank}[1]{\benchname{Bank}{}{#1}}
\newcommand{\constructor}{\benchname{Constructor}{}{}}
\newcommand{\bidding}[1]{\benchname{Bidding}{}{#1}}
\newcommand{\iqueue}{\benchname{IQueue}{}{}}
\newcommand{\keypad}{\benchname{Keypad}{}{}}
\newcommand{\simplequeue}{\benchname{SimpleQueue}{}{}}
\newcommand{\emmaba}{\benchname{EMMABA}{}{}}
\newcommand{\lazylist}{\benchname{LazyList}{}{}}
\newcommand{\cmsanyni}[2]{\benchname{CMSany}{ni}{#1\!\times\!#2}}
\newcommand{\cmsanygni}[2]{\benchname{CMSany}{gni}{#1\!\times\!#2}}
\newcommand{\cmsndetni}[2]{\benchname{CMSndet}{ni}{#1\!\times\!#2}}
\newcommand{\cmsndetgni}[2]{\benchname{CMSndet}{gni}{#1\!\times\!#2}}
\newcommand{\cmsmaxni}[2]{\benchname{CMSmax}{ni}{#1\!\times\!#2}}
\newcommand{\cmsmaxgni}[2]{\benchname{CMSmax}{gni}{#1\!\times\!#2}}
\newcommand{\isolrcser}[2]{\benchname{Isol}{rcser}{#1\!\times\!#2}}
\newcommand{\isolserrc}[2]{\benchname{Isol}{serrc}{#1\!\times\!#2}}
\newcommand{\cniany}[1]{\benchname{CNI}{any}{#1}}
\newcommand{\cnilte}[1]{\benchname{CNI}{lte}{#1}}
\newcommand{\selfstablf}[1]{\benchname{Selfstab}{lf}{#1}}
\newcommand{\selfstabgf}[1]{\benchname{Selfstab}{gf}{#1}}
\newcommand{\robustcrit}[1]{\benchname{Robust}{crit}{#1}}
\newcommand{\robustncrit}[1]{\benchname{Robust}{ncrit}{#1}}
\newcommand{\ctrlspecone}[2]{\benchname{Ctrl}{spec1}{#1\!\times\!#2}}
\newcommand{\ctrlspectwo}[2]{\benchname{Ctrl}{spec2}{#1\!\times\!#2}}
\newcommand{\mutantdk}[1]{Mutant#1$_{\text{\textsc{dk}}}$}
\newcommand{\mutantpk}[1]{Mutant#1$_{\text{\textsc{pk}}}$}

\newcommand{\Lang}{\mathcal{L}}
\newcommand{\A}{\mathcal{A}}
\newcommand{\B}{\mathcal{B}}

\newcommand{\LTLg}{\mathsf{G}}
\newcommand{\LTLf}{\mathsf{F}}

\newcommand{\na}{--}

\definecolor{provOld}{HTML}{E6E6E6}
\definecolor{provTosem}{HTML}{CFE2F3}
\definecolor{provNew}{HTML}{FCE5CD}
\definecolor{provEdgeT}{HTML}{3D85C6}
\definecolor{provEdge}{HTML}{E69138}

\newcommand{\nrows}{120}
\newcommand{\nrowsHyperQB}{47}
\newcommand{\nrowsHyperSMV}{25}
\newcommand{\nrowsHyperLasso}{48}
\newcommand{\timeoutsecs}{1000}

\newcommand{\solvedOmegaHyperHyperQB}{47}

\newcommand{\solvedOmegaHyperHyperSMV}{24}

\newcommand{\solvedOmegaHyperHyperLasso}{42}

\begin{document}

\title{\tool: Explicit-State HyperLTL Model Checking on an \texorpdfstring{$\omega$}{omega}-Automaton Diet}
\titlerunning{\tool}

\author{Hugo Pacheco\inst{1}\orcidID{0000-0003-0720-7744} }
\authorrunning{H.~Pacheco}
\institute{University of Minho \& INESC TEC, Portugal\\
\email{hpacheco@di.uminho.pt}}

\maketitle

\begin{abstract}
HyperLTL extends LTL with quantification over traces. \autohyper{} is the de facto explicit-state model
checker for HyperLTL, complete for an arbitrary quantifier prefix: it eliminates quantifiers
through products of the formula's $\omega$-automaton with the systems, complementing on quantifier alternations.
However, it suffers from state explosion, as it fully materializes input systems and intermediate automata.
We present \tool, an overhaul of \autohyper{} that keeps its algorithm but decomposes and reduces the size of the processed automata, and integrates the \kofola{} $\omega$-automaton complementation tool.
We compare \tool{} with existing explicit-state and symbolic HyperLTL model checkers, on \nrows{} benchmarks from their published suites.
\tool{} improves the performance of explicit-state HyperLTL model checking by up to orders of magnitude; bounded symbolic tools still reach further on violated instances with short counterexamples, as expected.
\end{abstract}

\section{Introduction}\label{sec:intro}

Hyperproperties~\cite{ClarksonS10} are specifications that relate multiple traces. They capture requirements that no property of single traces can express, such as information-flow policies and planning or synthesis tasks.
HyperLTL~\cite{ClarksonFKMRS14} extends Linear Temporal Logic (LTL) with explicit quantifiers over traces to specify general hyperproperties. Its model checking problem is decidable for finite-state systems, with a cost that grows nonelementarily in the number of quantifier alternations~\cite{FinkbeinerRS15}.

\autohyper~\cite{BeutnerF23,beutner2025autohyper} is the only complete model checker for arbitrary HyperLTL properties over arbitrary reactive systems. It converts each model into an explicit-state system, translates the quantifier-free body of the property into a $\omega$-automaton, and eliminates the quantifiers by computing products of the their systems with the automaton, complementing at every quantifier alternation.

Symbolic bounded model checkers, such as \hyperqb~\cite{HsuSB21,hyperqb2} and \hyperlasso~\cite{hyperlasso}, trade completeness for scale by encoding all traces up to a maximum bound into Quantified Boolean Formulas (QBF). This often finds violations of within that bound, but cannot in general establish that a hyperproperty holds.

Many relevant hyperproperties, and in fact most published benchmarks, have at most one alternation. As an example, consider the non-interference property of Smith and Volpano~\cite{SmithV98}, illustrated with a multi-threaded program that compares an internal variable \textit{mask} with a secret \textit{pin}, and where \textit{mask} also influences a public \textit{result}.
The property $\varphi_{\text{\textsc{CNI}}}$ states that, for every trace $\pi_1$ under a fair scheduler (one that runs every \textit{thread} of the set $\mathcal{T}$ infinitely often), there must be a trace $\pi_2$ with a different secret \textit{pin} that ends with the same public \textit{result}:
\begin{equation*}\small
\forall \pi_1.\,\exists \pi_2.\,\textstyle(\bigwedge_{t \in \mathcal{T}} \LTLg\LTLf\, \mathit{thread}_{\pi_1} {=} t) \rightarrow \,\neq\!{\mathit{pin}_{\pi_{\{1,2\}}}} \wedge \,\LTLf(\mathit{mask}_{\pi_{\{1,2\}}} {=} \, 0 \,\wedge =\!{\mathit{result}_{\pi_{\{1,2\}}}} )
\end{equation*}
\autohyper{} builds the product of the automaton from the body of $\varphi_{\text{\textsc{CNI}}}$ with the system for $\pi_2$, which yields an automaton $\B$ over $\pi_1$ accepting exactly the traces for which some $\pi_2$ exists; it then complements $\B$ and checks that it does not accept any trace of the system for $\pi_1$. The nonelementary cost of the logic is paid once, for the complement, which is exponential in the size of $\B$.

Such benchmarks fall into the most favourable class of true hyperproperties\footnote{Standard model checkers can handle alternation-free properties via self-composition.}, yet two things still confine \autohyper{} to small scopes.
The first is that every system is first enumerated into explicit states, and intermediate steps hand a fully materialised automata to external tools. This is to some extent intrinsic, whereas symbolic approaches avoid precisely such enumeration.
The second is that one alternation already builds the product of several systems, and complements an automaton built from them. \autohyper{} can trade the last complementation with calling an inclusion checker, which may search more lazily, but both operands of the inclusion must be fully built beforehand.

\subsection{Contributions}\label{sec:contributions}

This paper presents \tool{}, a large rework and extension of \autohyper{} that preserves its core algorithm and adds the following new features:
\begin{itemize}
\item \emph{Declarative models} (\Cref{sec:inputmodels}): the translation of \nusmv{} models reads \texttt{INIT}, \texttt{INVAR} and \texttt{TRANS} clauses, and keeps only states within infinite paths.
\item \emph{Symbolic edges} (\Cref{sec:inputmodels}): frozen and free variables of a model stay on the edges of an otherwise fully-explicit system, and are enumerate when needed.
\item \emph{Formula decomposition} (\Cref{sec:inputformula}): the formula is split into parts, and the pipeline is lifted to work over partitioned formulas and automata.
\item \emph{Product quotients} (\Cref{sec:elimination}): \autohyper{} quotients input systems via bisimulation; \tool{} also quotients every product automaton.
\item \emph{Support for \kofola{}} (\Cref{sec:final}): \tool{} delegates every complementation and inclusion check to \kofola{}~\cite{AlexajHHLLM26}, a tool that decomposes an automaton and complements it modularly. We extend \kofola{} inclusion checking to read automata parts directly and to return counterexamples.
\end{itemize}

We also perform an evaluation (\Cref{sec:eval}) of \tool{} on \nrows{} benchmarks from three published suites against existing state-of-the-art tools.

\section{Related Work}\label{sec:related}

Considerable work has addressed the verification of hyperproperties expressed in general hyperlogics, in particular HyperLTL~\cite{Finkbeiner23}. We focus on model checking; hyperproperties over programs have also been verified deductively~\cite{DardinierLM24,Beutner24} or through constrained Horn clauses~\cite{ItzhakySV24}, and refuted via symbolic execution~\cite{CorrensonNFW24}.

\paragraph{Explicit-state approaches.} \cite{ClarksonFKMRS14} showed that HyperLTL model checking is decidable. \mchyper~\cite{FinkbeinerRS15} was the first model checker to handle models of some complexity: it reduces alternation-free formulas to a self-composition, checked by a hardware model checker. For $\forall^*\exists^*$ formulas, a strategy allows to choose each step of the existential traces from the universal traces seen so far; it may be provided by the user~\cite{FinkbeinerRS15} or synthesised by solving a game~\cite{CoenenFST19}. Strategy-based checking needs prophecy variables to become complete~\cite{BeutnerF22csf}, and extends to asynchronous hyperlogics~\cite{BeutnerF22cav,BeutnerF25fmsd,BeutnerF25oopsla}.
\autohyper~\cite{BeutnerF23} decides the full logic through automata complementation, and its evaluation suggests that it outperforms strategy-based checking on HyperLTL. Its current version, which \tool{} builds on, supports also propositional quantification~\cite{BeutnerF23lpar} and HyperLTL modulo theories~\cite{beutner2025autohyper}.

\paragraph{Symbolic approaches.} \hyperqb~\cite{HsuSB21} encodes arbitrary HyperLTL properties into QBF by unrolling systems up to a bound. However, it is only conclusive in one direction, depending on the finite-trace semantics approximation that is chosen, and in both directions only for terminating systems. Loop conditions~\cite{HsuSSB23} handle lasso-shaped paths for $\forall\exists\,\LTLg \varphi$ and $\exists\forall\,\LTLg \varphi$ properties, but enumerate the states of the universally-quantified system.
Model checking the asynchronous logic A-HLTL~\cite{BaumeisterCBFS21} is undecidable in general; its bounded procedure~\cite{HsuBFS23} is conclusive only for terminating systems. The latest version of \hyperqb~\cite{hyperqb2} integrates all these advances and adds an SMT encoding.
\hyperlasso~\cite{hyperlasso} is a bounded model checker for refuting
$\forall^*\exists^*$ hyperproperties over non-terminating systems: an SMT-based synthesizer proposes lasso-shaped counterexamples, and an LTL model checker confirms them; proving such properties remains inconclusive.

\paragraph{Hybrid approaches.} \tool{} avoids enumerating a model's frozen and free variables, by keeping them on the edges of its explicit system as symbolic guards, so that one explicit state stands for all their valuations.
It performs a set-based reduction of the explicit state space, similar to what the \divine{} model checker applies to programs~\cite{BauchHB16}.
Hybrid model checking~\cite{sebastiani2011hybrid} considers a fully symbolic system against an explicit property. Both approaches targeted only plain LTL.

\paragraph{Input formats.} Existing hyper model checkers take HyperLTL formulas written in a largely common syntax and read models either in low-level formats or an imperative subset of the SMV language.
\hypersmv~\cite{hypersmv} is a pre-processor that supports declarative SMV models; it is demonstrated using an explicit-state backend (a modified version of \autohyper{}) and a dedicated symbolic backend built on QBF solvers.
\autohyper{} further requires the formula to mark its \emph{atomic propositions} (APs): the non-temporal sub-formulas (such as $x_{\pi_1} \!= x_{\pi_2}$) that fix the alphabet of every automaton and will be read as single Boolean letters.\footnote{In \autohyper{}, APs are indeed mandatory for some sub-expressions such as comparisons, which is enforced at the level of the parser.} To free users from such obligation, \hypersmv{} automatically infers and manages APs.
\tool{} extends the \autohyper{} version of \hypersmv{} and integrates these \hypersmv{} features.

\paragraph{Automata backends.} \autohyper~\cite{BeutnerF23} relies on \spot~\cite{Duret-LutzRCRAS22} for LTL translation, automaton reduction and complementation, and calls \spot, \rabit~\cite{AbdullaCCHHMV11}, \bait~\cite{DoveriGPR21}, \forklift~\cite{DoveriGM22}, \forq{} (\spot{}'s reimplementation of \forklift{}) and \roll~\cite{LiSTCX19} for inclusion checking.
\tool{} preserves all those backends and provides novel support for \kofola~\cite{HavlenaLLST23,AlexajHHLLM26}. Its modular construction partitions an automaton into its strongly connected components (SCCs), groups them by type,
and complements each with a dedicated construction.

\section{Automata-based HyperLTL Model Checking}\label{sec:background}

Automata-based model checking~\cite{ClarkeGP99} turns a temporal property $\psi$ into an $\omega$-automaton that accepts the (infinite) words satisfying $\psi$; the automaton for $\neg\psi$ then accepts the counterexamples to $\psi$, and its product with the system accepts the counterexamples that are executions of the system, so verification of $\psi$ over the system reduces to the emptiness of that product.
LTL is translated into nondeterministic Büchi automata (NBAs), which accept a word when some run on it visits an accepting state infinitely often, and are closed under the operations that model checking needs (product and complementation).
A non-empty NBA always accepts a \emph{lasso}, a finite prefix followed by a finite cycle repeated forever.

HyperLTL lifts this to multiple traces.
Consider a formula in prenex normal form $\varphi = Q_1\pi_1.\cdots Q_n\pi_n.\,\psi$, with $Q_i\in\{\forall,\exists\}$, where we quantify over the traces of one or more systems; consecutive quantifiers of the same kind form a \emph{block}. Its body $\psi$ is an LTL formula over the variables of named traces, for example $\LTLg (x_{\pi_1} = x_{\pi_2})$.
The automaton for the body reads the variables of all quantified traces at once, so a word is a tuple of traces, one per trace variable.
For automata-based HyperLTL model checking, each block has to be eliminated in turn, and each quantifier alternation will cost a complementation.

\begin{algorithm}[t]
\caption{\autohyper's model checking of $\varphi = Q_1\pi_1.\cdots Q_n\pi_n.\,\psi$}\label{alg:autohyper}
\begin{algorithmic}[1]
\State $(\B,\mathit{neg}) \gets (\mathrm{LTLtoNBA}(\textbf{if } Q_n=\exists \textbf{ then } \psi \textbf{ else } \neg\psi),\ Q_n\neq\exists)$ \Comment{convert formula}
\For{$i = n$ \textbf{downto} $1$}
  \If{$\mathit{neg} \neq (Q_i=\forall)$} $(\B,\mathit{neg}) \gets (\overline{\B},\neg\mathit{neg})$ \Comment{at a quantifier alternation}
  \EndIf
  \State $\B \gets \A_{\pi_i}\otimes\B$ \Comment{eliminate quantifier}
\EndFor
\State \Return $\mathit{neg} \Leftrightarrow \Lang(\B)=\emptyset$ \Comment{final check}
\end{algorithmic}
\end{algorithm}

\paragraph{\autohyper's algorithm~\cite{BeutnerF23}.} Each trace variable $\pi$ ranges over the traces of a system, read as an NBA $\A_\pi$ whose words are the system's traces (unless a model states otherwise, every state of $\A_\pi$ is accepting). For an NBA $\B$ over the traces of $\pi_1,\dots,\pi_i$, then $\A_{\pi_i}\otimes\B$ is the product of $\A_{\pi_i}$ with $\B$ with the trace of $\pi_i$ projected away: it accepts the traces for $\pi_1,\dots,\pi_{i-1}$ that some trace of $\A_{\pi_i}$ extends to a word of $\B$. $\overline{\B}$ is the complement of $\B$. The language $\Lang(\B)$ of an automaton $\B$ denotes the set of words it accepts.

For an input formula in prenex normal form, \Cref{alg:autohyper} iteratively eliminates the innermost quantifiers.
It carries two variables: an automaton $\B$ and a flag $\mathit{neg}$. After $\pi_i$ is eliminated, $\B$ accepts the traces for $\pi_1,\dots,\pi_{i-1}$ that satisfy $Q_i\pi_i.\cdots Q_n\pi_n.\,\psi$, or its negation if $\mathit{neg}$ holds. Every quantifier is eliminated by a product with its system, which projects the trace away and behaves as an existential; $\mathit{neg}$ must hold exactly for universal quantifiers, where $\forall\pi.\,\varphi' = \neg\exists\pi.\,\neg\varphi'$ is applied. Quantifier alternations require a complementation and flipping $\mathit{neg}$. After the last iteration, $\B$ reads no trace and the verdict is given by an emptiness check: under $\mathit{neg}$, $\B$ stands for the negation of $\varphi$, so $\varphi$ holds iff $\B$ is empty, as in classical LTL model checking; otherwise $\varphi$ holds iff $\B$ is not empty.

\autohyper{} applies two additional refinements not directly shown in~\Cref{alg:autohyper}.
First, the algorithm eliminates per block (where all quantifiers share the same value of $\mathit{neg}$) instead of per quantifier; this simply amounts to performing the product with the self-composition $\A = \A_{\pi_1}\times\cdots\times\A_{\pi_k}$ of that blocks' systems.
Second, when the outermost block is universal, the last iteration complements $\B$, takes its product with the self-composition $\A$ of that block's systems, and checks emptiness. \autohyper{}'s \emph{inclusion mode} instead stops before that block and checks $\Lang(\A)\subseteq\Lang(\B)$, hiding the last and largest complementation inside the inclusion checker: note that $\Lang(\A)\subseteq\Lang(\B) \Leftrightarrow \Lang(\A\otimes\overline{\B})=\emptyset$, but inclusion checkers can explore the product $\A\otimes\overline{\B}$ on the fly without building the whole $\overline{\B}$.

\paragraph{Atomic propositions.} The formula's NBA does not read the systems' variables directly: its \emph{letters} are valuations of \emph{atomic propositions} (APs). Its \emph{alphabet} is the set $2^{AP}$, and each of its transitions carries a \emph{label}, a Boolean formula over the APs that stands for the letters satisfying it. In \autohyper{}, the user marks each AP explicitly in the formula as a Boolean expression over the variables of named traces, which may relate several traces, as in $x_{\pi_1} = x_{\pi_2}$. $\mathrm{LTLtoNBA}$ translates the formula over these APs, and the product later decides an AP's truth from the systems' states.
The choice of APs matters because \spot{}'s translation grows quickly with their number. Moreover, APs change during~\Cref{alg:autohyper}: a product substitutes the values of the trace it eliminates into every AP and simplifies it~\cite{beutner2025autohyper}; an AP that becomes a constant is dropped, and one that still reads the remaining traces survives as a \emph{residual AP} of the new automaton, created lazily as the product reaches it.
For instance, once $\pi_2$ is eliminated, $x_{\pi_1} = x_{\pi_2}$ becomes one AP $x_{\pi_1} = v$ for every value $v$ that $x$ takes in a state of $\pi_2$'s system.

\section{\tool: Tool Overview}\label{sec:architecture}

\tool{} is an explicit-state model checker for HyperLTL over \nusmv{} models. It is developed as an extension of \autohyper{}. Its command line takes one \nusmv{} model per trace (or one for all traces) and a formula:\footnote{\autohyper{} also supports reading lower-level formats such as explicit-state systems or boolean programs. \tool{} keeps those parsers intact.}
\begin{center}\small
\verb|OmegaHyper --nusmv cni_any_3.smv cni_3.ah --witness|
\end{center}
For a refuted outermost-$\forall$ or a proven outermost-$\exists$ property, \flag{witness} prints one lasso per model of the outermost block.
All extensions shown below are on by default, but each has a command-line switch that turns it off.

\begin{figure}[t]
\begin{minipage}[t]{0.42\textwidth}
\scriptsize\begin{alltt}
\exLine{1}MODULE main
\exLine{2}VAR \exRigid{pin : word[3];}    mask : word[3];
\exLine{3}    result : word[3]; \exInput{thread : 0..2;}
\exLine{4}\exDecl{INIT pin > mask}
\exLine{5}ASSIGN
\exLine{6}  \exRigid{next(pin)    := pin;}
\exLine{7}  next(mask)   := case
\exLine{8}   thread=2 & ... : mask >> 1;
\exLine{9}   TRUE           : mask; esac;
\exLine{10}  init(result) := 0;
\exLine{11}  next(result) := case
\exLine{12}   thread=0 & ... : result | mask;
\exLine{13}   thread=1 & ... : result & !mask;
\exLine{14}   TRUE           : result; esac;
\exLine{15}  ...
\end{alltt}
\end{minipage}\hfill
\begin{minipage}[t]{0.56\textwidth}
\centering
\begin{tikzpicture}[baseline={([yshift=-6pt]current bounding box.north)}, x=23mm, y=-13mm,
  box/.style={draw=black!70, inner sep=2pt, font=\scriptsize\ttfamily},
  ed/.style={-{Stealth[length=1.4mm]}, black!70},
  dots/.style={fill=white, inner sep=0.6pt, font=\scriptsize, text=black!70}]
\newcommand{\st}[4]{\setlength{\tabcolsep}{2pt}\begin{tabular}{@{}c|c@{}}#1&#2\\\hline#3&#4\end{tabular}}
\node[box] (leg) at (0,0) {\setlength{\tabcolsep}{2pt}\begin{tabular}{@{}c|c@{}}pin&thread\\\hline mask&result\end{tabular}};
\node[box] (s0) at (1,0) {\st{001}{0}{010}{000}};
\node[box] (s1) at (2,0) {\st{\g{001}}{2}{\g{010}}{010}};
\node[box] (s2) at (2,1) {\st{\g{001}}{0}{001}{\g{010}}};
\node[box] (s3) at (1,1) {\st{\g{001}}{1}{\g{001}}{011}};
\node[box] (s4) at (0,1) {\st{\g{001}}{2}{\g{001}}{010}};
\node[box] (l0) at (0,2) {\st{\g{001}}{0}{000}{\g{010}}};
\node[box] (l1) at (1,2) {\st{\g{001}}{1}{\g{000}}{\g{010}}};
\node[box] (l2) at (2,2) {\st{\g{001}}{2}{\g{000}}{\g{010}}};
\draw[ed] ([xshift=-7.5mm]s0.west) -- node[dots] {$\cdots$} (s0);
\draw[ed] (s0) -- node[dots] {$\cdots$} (s1);
\draw[ed] (s1) -- node[dots] {\rotatebox{90}{$\cdots$}} (s2);
\draw[ed] (s2) -- node[dots] {$\cdots$} (s3);
\draw[ed] (s3) -- node[dots] {$\cdots$} (s4);
\draw[ed] (s4) -- (l0);
\draw[ed] (l0) -- (l1);
\draw[ed] (l1) -- (l2);
\draw[ed] (l2.south) -- ++(0,-3.2mm) -| (l0.south);
\end{tikzpicture}
\end{minipage}
\caption{The CNI benchmark, abridged to the variables the property reads.
Left: \nusmv{} model; \exRigid{orange} marks a frozen variable, \exInput{blue} a free variable, and \exDecl{green} the declarative clause of \cnilte{3}.
Right: depiction of the lasso counterexample for \cniany{3}; the legend at the top left names the cells of a state; dots ($\cdots$) ignore steps that change neither \texttt{mask} nor \texttt{result}; \g{grey} values did not change since the previous state.}
\label{fig:example}
\end{figure}
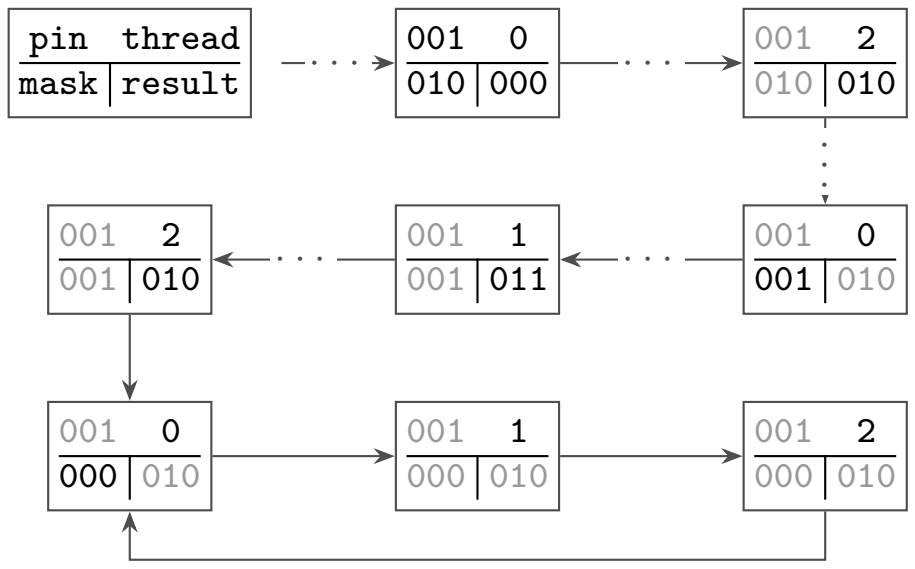

\paragraph{Running example.}
Consider the $\varphi_{\text{\textsc{CNI}}}$ property of~\Cref{sec:intro}, which is the CNI benchmark of~\cite{hyperlasso}; the NI benchmark of~\cite{hyperqb2} models a simpler variant without fairness.
The model of \Cref{fig:example} defines three \texttt{thread}s, of which one runs at each step.
The main thread (\texttt{thread=2}) checks $\mathtt{pin} = \mathtt{mask}$ and triggers the other two: \texttt{thread=0} sets the \texttt{result} bits that are set in \texttt{mask} (line 12) and \texttt{thread=1} clears them (line 13).
If the equality holds \texttt{thread=0} runs first, and \texttt{thread=1} otherwise; the second undoes the effects of the first on \texttt{result}.
For scope 3, the number of bits, the benchmark defines two models: \cniany{3}, where the initial \texttt{mask} is arbitrary; and \cnilte{3}, which constraints it with the clause at line 4.

On \cniany{3}, $\varphi_{\text{\textsc{CNI}}}$ is false and \tool{} returns the $\pi_1$ counterexample from~\Cref{fig:example}.
Fairness is met: the 3-state loop schedules every thread infinitely often.
As \texttt{mask} decreases, it becomes equal to \texttt{pin} (third state), and the equality check leaks \texttt{pin} into \texttt{result}.
Since no other \texttt{pin} halts with $\mathtt{result} = 010$, no $\pi_2$ exists: \texttt{mask} always takes the value $001$ last before the program halts, hence in every halted trace the least significant bit of \texttt{result} is 0 exactly when $\mathtt{pin} = 001$, and a trace with another \texttt{pin} would halt with that bit set.

On \cnilte{3}, $\varphi_{\text{\textsc{CNI}}}$ is true: the equality never holds, since $\mathtt{pin} > \mathtt{mask}$ at the start and \texttt{mask} only decreases; the final \texttt{result} is then a function of the initial \texttt{mask} alone, independent of \texttt{pin}, and a $\pi_2$ always exists.

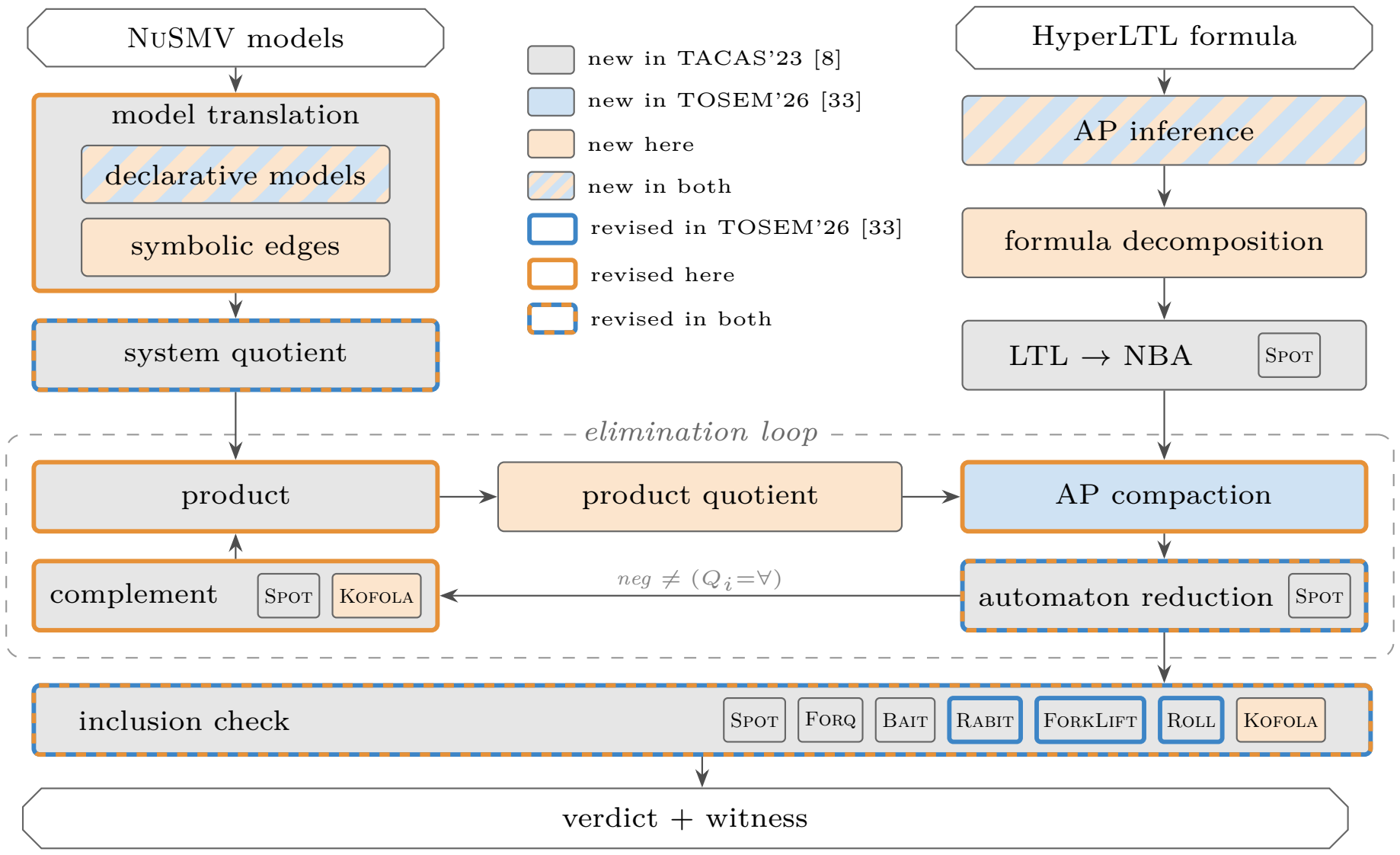
\begin{figure}[!t]
\centering
\begin{tikzpicture}[
  font=\scriptsize,
  node distance=2.2mm and 3mm,
  box/.style={draw=black!60, rounded corners=1.5pt, align=center, inner sep=2.5pt,
              minimum height=6mm, text width=#1},
  box/.default=33.3mm,
  old/.style={box, fill=provOld},
  tosem/.style={box, fill=provTosem},
  new/.style={box, fill=provNew},
  bands/.style={fill=provTosem, path picture={
      \foreach \k in {-40,...,40} {
        \fill[provNew] ([xshift=\k*2*#1]path picture bounding box.south west) -- ++(#1,0) -- ++(30mm,30mm) -- ++(-#1,0) -- cycle;
      }}},
  newTO/.style={box, bands=6pt},
  revT/.style={draw=provEdgeT, line width=1.1pt},
  revO/.style={draw=provEdge, line width=1.1pt},
  revTO/.style={draw=provEdgeT, line width=1.1pt,
                postaction={draw=provEdge, line width=1.1pt, dash pattern=on 2.5pt off 2.5pt}},
  sub/.style={draw=black!60, rounded corners=1pt, inner sep=1.6pt, minimum height=3.8mm, font=\tiny},
  group/.style={box, fill=provOld, minimum height=11.5mm},
  inner/.style={text width=25mm, minimum height=5mm},
  io/.style={draw=black!60, fill=white, align=center, inner sep=2pt, minimum height=5mm,
             text width=#1, chamfered rectangle, chamfered rectangle xsep=2pt},
  io/.default=33.3mm,
  flow/.style={-{Stealth[length=1.6mm]}, black!70},
  lane/.style={font=\scriptsize\itshape, text=black!60},
  swatch/.style={rounded corners=1pt, minimum width=4mm, minimum height=2.4mm, inner sep=0pt},
  legend/.style={inner sep=0pt, anchor=west, font=\tiny},
  cond/.style={font=\tiny, text=black!50, inner sep=1pt},
]
\node[io]                         (model) {\nusmv{} models};
\node[old, revO, below=of model, minimum height=17mm] (decl) {};
\node[anchor=north, inner sep=2.6pt] at (decl.north) {model translation};
\node[new, inner, anchor=south] (feat) at ([yshift=1.4mm]decl.south) {symbolic edges};
\node[newTO, inner, above=1.2mm of feat] (dm) {declarative models};
\node[old, revTO, below=of decl]  (bisim) {system quotient};

\node[old, revO, below=5.9mm of bisim] (prod) {product};
\node[new, right=5mm of prod]         (pred) {product quotient};
\node[tosem, revO, right=5mm of pred] (apc0) {AP compaction};
\node[old, revO, below=of prod] (comp) {};
\node[anchor=west, inner sep=0pt] at ([xshift=1.5mm]comp.west) {complement};
\matrix[column sep=1mm, anchor=east, inner sep=0pt] at ([xshift=-1.5mm]comp.east) {
  \node[sub, fill=provOld] {\spot}; & \node[sub, fill=provNew] {\kofola}; \\};
\node[old, revTO] at (apc0 |- comp) (aut) {};
\node[anchor=west, inner sep=0pt] at ([xshift=1.5mm]aut.west) {automaton reduction};
\node[sub, fill=provOld, anchor=east] at ([xshift=-1.5mm]aut.east) {\spot};
\begin{scope}[on background layer]
  \node[draw=black!40, dashed, rounded corners, fit=(prod)(pred)(apc0)(comp)(aut), inner sep=2.2mm,
        label={[lane, anchor=center, fill=white, inner sep=1pt]north:elimination loop}] (elim) {};
\end{scope}

\node[io] at (apc0 |- model)      (phi)   {HyperLTL formula};
\node[newTO, below=of phi]        (merge) {AP inference};
\node[old] at (bisim -| phi)      (ltl)   {};
\node[anchor=west, inner sep=0pt] at ([xshift=4mm]ltl.west) {LTL $\to$ NBA};
\node[sub, fill=provOld, anchor=east] at ([xshift=-4mm]ltl.east) {\spot};
\node[new] at ($(merge)!0.5!(ltl)$) (restr) {formula decomposition};

\path let \p1=(prod.west), \p2=(comp.south), \p3=(aut.east) in
  node[old, revTO, anchor=north west, text width=\x3-\x1-5pt] (incl) at (\x1,\y2-4.4mm) {};
\node[anchor=west, inner sep=0pt] at ([xshift=4mm]incl.west) {inclusion check};
\matrix[column sep=1mm, anchor=east, inner sep=0pt] at ([xshift=-4mm]incl.east) {
  \node[sub, fill=provOld] {\spot}; & \node[sub, fill=provOld] {\forq}; & \node[sub, fill=provOld] {\bait}; &
  \node[sub, fill=provOld, revT] {\rabit}; & \node[sub, fill=provOld, revT] {\forklift}; &
  \node[sub, fill=provOld, revT] {\roll}; & \node[sub, fill=provNew] {\kofola}; \\};
\path let \p1=(incl.west), \p2=(incl.south), \p3=(incl.east) in
  node[io, anchor=north west, text width=\x3-\x1-12pt] (out) at (\x1,\y2-3.5mm) {verdict $+$ witness};

\draw[flow] (model) -- (decl);
\draw[flow] (decl) -- (bisim);
\draw[flow] (phi) -- (merge);
\draw[flow] (merge) -- (restr);
\draw[flow] (restr) -- (ltl);
\draw[flow] (bisim) -- (prod);
\draw[flow] (ltl) -- (apc0);
\draw[flow] (prod) -- (pred);
\draw[flow] (pred) -- (apc0);
\draw[flow] (apc0) -- (aut);
\draw[flow] (aut) -- node[cond, above] {$\mathit{neg}\neq(Q_i{=}\forall)$} (comp);
\draw[flow] (comp) -- (prod);
\draw[flow] (aut.south) -- (aut.south |- incl.north);
\draw[flow] (incl.south) -- (incl.south |- out.north);

\coordinate (lg) at ($(pred |- model.north)!0.42!(pred |- elim.north)$);
\node[swatch, draw=black!60, fill=provOld, anchor=north west] (l1) at ([xshift=-15mm, yshift=12.3mm]lg) {};
\node[legend, right=1mm of l1] {new in TACAS'23~\cite{BeutnerF23}};
\node[swatch, draw=black!60, fill=provTosem, below=1.1mm of l1.south west, anchor=north west] (l2) {};
\node[legend, right=1mm of l2] {new in TOSEM'26~\cite{hypersmv}};
\node[swatch, draw=black!60, fill=provNew, below=1.1mm of l2.south west, anchor=north west] (l3) {};
\node[legend, right=1mm of l3] {new here};
\node[swatch, draw=black!60, bands=2.5pt, below=1.1mm of l3.south west, anchor=north west] (l4) {};
\node[legend, right=1mm of l4] {new in both};
\node[swatch, fill=white, revT, below=1.1mm of l4.south west, anchor=north west] (l5) {};
\node[legend, right=1mm of l5] {revised in TOSEM'26~\cite{hypersmv}};
\node[swatch, fill=white, revO, below=1.1mm of l5.south west, anchor=north west] (l6) {};
\node[legend, right=1mm of l6] {revised here};
\node[swatch, fill=white, revTO, below=1.1mm of l6.south west, anchor=north west] (l7) {};
\node[legend, right=1mm of l7] {revised in both};
\end{tikzpicture}
\caption{The \tool{} architecture. The fill of each step gives origin (original \autohyper{}~\cite{BeutnerF23},~\cite{hypersmv} or this paper; stripes denote \hypersmv{}-only features reimplemented in this paper) and border signals revisions. Small boxes denote invoked tools.}
\label{fig:architecture}
\end{figure}

\paragraph{Architecture.}
\tool{} builds on the \autohyper{} version of~\cite{hypersmv}.
\Cref{fig:architecture} shows its pipeline.
Each \nusmv{} model, imperative or \emph{declarative}, is translated into an explicit-state system, whose frozen and free variables stay as \emph{symbolic edges}, then \emph{quotiented} by bisimulation. The formula's APs are \emph{inferred}, its body is \emph{decomposed} into parts, each being \emph{translated} into an automaton.
Each iteration of the \emph{elimination loop} works over the partitioned automaton: it \emph{compacts} its APs, \emph{reduces} it, and \emph{complements} it at quantifier alternations; the \emph{product} with the block's systems is then \emph{quotiented}.
The last block is decided by an \emph{inclusion check}, whose counterexample becomes a witness over the models' variables.

\paragraph{Implementation.} \tool{} is written in $\sim$15k lines of F\# (.NET~9). As external tools, it requires \spot{} for LTL translation and reduction, and our own fork of \kofola{} for complementation and inclusion.
Our \kofola{} extension amounts to 500 lines of C++.
\spot, \rabit, \bait, \forklift, \forq{} and \roll{} remain available as optional inclusion backends.

\section{Inside \tool}\label{sec:techniques}

Explicit-state HyperLTL model checking pays for the size of its inputs three times: in the states of the systems it enumerates, in the alphabets of the automata it translates and complements, and in the products it computes. This section follows the pipeline of~\Cref{fig:architecture}, and explains precisely what \tool{} does at each step to keep these sizes in check. Every extension is exact and preserves the verdict; a command-line switch turns it off, changing only overall cost.

\subsection{Input Systems}\label{sec:inputmodels}

\paragraph{Declarative models.}

\autohyper{} reads only imperative SMV models, with \texttt{INIT} and \texttt{ASSIGN} clauses per variable, which can be enumerated almost directly into an explicit-state system.
\tool{} also reads declarative models, whose \texttt{INIT}, \texttt{INVAR} and \texttt{TRANS} clauses are global, and where the successors of a state are the valuations of all variables that satisfy every clause jointly.
We reimplement the general translation procedure of \hypersmv{}: each clause is compiled into a decision diagram in which every variable ranges over its own finite domain, and the successors of a state are enumerated by a search over these diagrams.

The translation of well-formed declarative SMV model may leave states without successors, which lie on no infinite path and are removed.
Note that \autohyper{} needed no such step, as in a well-formed imperative SMV model every variable always has a next value.
As~\hypersmv{}, we also identify and move invariants from the formula into the models, so that model translation never enumerates states that violate such invariants.

In our example, the clause \texttt{INIT pin > mask} of \cnilte{3} (\Cref{fig:example}) is declarative, and \autohyper{} rejects it. An imperative model can leave a variable's initial value open (an \texttt{INIT} with a set of values); but this clause relates two variables, and expressing it in imperative style would need a re-encoding with auxiliary variables.
Another option would be moving the constraint into the formula, at the cost of a needlessly larger system (which \cite{hyperlasso} did in its evaluation).

\paragraph{Symbolic edges.}
A variable is \emph{frozen} if it never changes and \emph{free} if no clause relates its values across steps.
Instead of enumerating such variables, which multiplies the state space by the number of their valuations, in \tool{} the states hold only the remaining, \emph{dynamic} variables, and every edge carries a guard: a decision diagram that says under which frozen/free values the step is possible.

An explicit-state system, as \autohyper{} reads it, is a state-labelled NBA: a state fixes every variable, hence the letter (the truth of every AP) that all its outgoing edges read. Labelling the edges instead with Boolean formulas over the APs, each standing for a set of letters, recognises the same languages, often with fewer states.
Free variables fit this form: a step chooses their value, so an edge's label stands for the letters of the values its guard allows.
Frozen variables do not: their value must stay the same along the whole run, which no single step's label can enforce, so a system with symbolic edges is not strictly an NBA.
In contrast, $\B$ must be an NBA as it is processed by automata tools; it will be the product's role to combine it with the systems back into a new NBA.

In our example, \texttt{pin} is frozen, \texttt{thread} is free, and \texttt{mask} and \texttt{result} are dynamic: for \cniany{3}, \texttt{mask} starts arbitrary but is not free since it is not chosen afresh at every step.
\cnilte{3} and \cniany{3} enumerate to 349 and 731 dynamic states, with 901 and 1934 edges whose guards range over the 8 values of the 3-bit \texttt{pin} and the 3 values of \texttt{thread}; without symbolic edges, these 24 valuations are enumerated into the states instead, and the systems have 3381 and 9342 explicit states (out of $349 \times 24$ and $731 \times 24$ combinations, not all reachable).

\paragraph{System quotient.}
Each system is then reduced to its bisimulation quotient by partition refinement~\cite{KanellakisS90}: the states start in equivalence classes given by what is observed of them, and a class is split until all of its states have transitions into the same classes.
In \autohyper{}, two states start in the same class when they agree on the value of every variable the formula reads on that system.
\cite{hypersmv} coarsened this to the truth of the formula APs that read only that system, plus the values read by APs over several systems.
\tool{} observes the latter by their residual APs instead, and compares the states' guards into each class as decision diagrams, so that merged states are bisimilar under every frozen/free valuation; the products, whose edges also carry labels, will reuse the same quotient.

In our example, the quotient takes each system of \cnilte{3} from 349 to 305 states, and each system of \cniany{3} from 731 to 606.

\subsection{Input Formula}\label{sec:inputformula}

\paragraph{AP inference.} The frontend of \hypersmv{} inferred APs before model checking, but only for formulas declaring none.
It followed two rules: every maximal non-temporal sub-expression over a single trace became one AP; and multi-trace subexpressions matching relational patterns between two variables were collapsed into one AP.
\tool{} accepts a formula with declared or undeclared APs, and infers APs inside the model checker: declared APs are never opened, but may be merged into new APs.
The multi-trace patterns of \hypersmv{} are generalized by a heuristic: a maximal non-temporal sub-expression over several traces is folded into one AP when the residual APs it is estimated to produce after elimination do not outnumber the APs it collapses.

In our example, the 21 APs of the input formula become 9: one per value of $\mathit{thread}_{\pi_1}$; one for each of $\mathit{mask}_{\pi_1} = 0$ and $\mathit{mask}_{\pi_2} = 0$, each aggregating three bits; one for $\mathit{pin}_{\pi_1} \neq \mathit{pin}_{\pi_2}$, aggregating three bit comparisons; and one per bit of $\mathit{result}_{\pi_1} = \mathit{result}_{\pi_2}$, which stay apart.

\paragraph{Formula decomposition.}
Throughout \tool{}'s algorithm, $\B$ is kept as a set of \emph{parts} $B_1,\dots,B_m$ together with a connective, $\B = \bigvee_i B_i$ or $\B = \bigwedge_i B_i$.
We have that
$\exists \vec\pi.\,\textstyle\bigvee_i B_i = \bigvee_i \exists \vec\pi.\,B_i$ and
$\forall \vec\pi.\,\textstyle\bigwedge_i B_i = \bigwedge_i \forall \vec\pi.\,B_i$, so each part quantifies its own copy of the innermost block's traces $\vec\pi$.
The connective is aligned with the $\mathit{neg}$ variable of~\Cref{alg:autohyper}, flipping at every quantifier alternation.

The first partitioning is made on the input formula.
For expressiveness, a model $M$ may carry its own \texttt{LTLSPEC}~$\varphi$, read as a restriction of its traces $\pi$ to those satisfying~$\varphi$, or fairness constraints (such as \texttt{JUSTICE}~$\varphi$) that behave the same way. Such restrictions constrain whole paths instead of states, and \tool{} moves them into the formula.
In the other direction, an invariant $\LTLg\,\phi_\pi$ in the body of the formula, where $\phi_\pi$ is non-temporal and reads only the trace $\pi \in M$, constraints only states and is moved from the formula into the model, where it is treated exactly as an \texttt{INVAR}~$\phi$ clause.
Both moves follow similar transformations found in \hypersmv{} and respect the prenex form of the formula: moving a part over $\pi$ across inner quantifiers inside relies on prenex equivalences such as $\varphi_\pi \rightarrow \exists\pi'.\,\psi = \exists\pi'.\,(\varphi_\pi \rightarrow \psi)$, which hold only when the crossed system pertaining to $\pi'$ has at least one trace. A quantifier dissolves over an empty system. \tool{} simplifies the formula before entering the loop, and ensures that every system that the loop meets is non-empty.

In our example, the body is an implication whose premise reads only $\pi_1$, so under the innermost $\exists \pi_2$ it splits into two parts: the three negated fairness conjuncts, which read no $\pi_2$, and the conclusion.

\paragraph{LTL $\to$ NBA.}
\tool{} finishes processing the formula by converting each formula part into a separate NBA.
When needed, the parts are combined into a single automaton, which is a linear operation, whereas \spot's translation grows superlinearly with the formula.
We delay the actual conversion of each formula part until it reaches a product, so that it can be negated syntactically (instead of being complemented) in case it reaches a quantifier alternation first.

In our example, the three premises stay as formulas, and \spot{} translates the conclusion into an NBA of 3 states, against 7 states for the whole body.

\subsection{Elimination Loop}\label{sec:elimination}

After processing the models and formula, each iteration of the loop in~\Cref{fig:architecture} eliminates the innermost quantifier block, running the following steps in order.

\paragraph{AP compaction.}
An AP that straddles a block reads an eliminated trace and a remaining one, and \autohyper{} leaves it as residual APs over the remaining trace alone, one per possible value in the eliminated trace.
\cite{hypersmv} compacted such APs just before the product: a comparison of two variables was rewritten over their value bits, so that it leaves logarithmically many residual APs.
\tool{} instead introduces \emph{defined APs}: each defined AP of the automaton is paired with a definition, a Boolean expression over input APs (original or inferred).
Compaction can thus run first in the loop: the reduction and the complement already work over the compact alphabet, while the product still reads the finer input APs.
Compaction comes in two halves.

The \emph{syntactic} half runs at every iteration and serves the product. It generalises the rewrite of~\cite{hypersmv} from comparisons between two variables to any straddling AP: its definition is the AP's own expression, with the parts over one trace kept as input APs and every multi-trace sub-expressions rewritten over the compared variables' value bits.
Following the rationale used for AP inference, the AP is only replaced with its definition when it would leave more residuals than the definition has inputs on the remaining traces.

The \emph{semantic} half is novel and runs only before a complement, the one step whose cost depends on the whole alphabet, letters no system can write included.
As formulas over the variables, many AP valuations never occur together (such as $x_{\pi} = 1$ and $x_{\pi} = 2$).
This half quotients the alphabet to keep only the letters the remaining systems can write, and groups those that satisfy the same labels of the automaton; groups are encoded in binary form, yielding a logarithmic encoding of the new defined APs, as the syntactic half encodes values.

In our example, only the syntactic half runs: the single alternation is decided by the final inclusion check, so there is no complement.
It keeps the 9 inferred APs: 4 straddle both traces ($\mathit{pin}_{\pi_1} \neq \mathit{pin}_{\pi_2}$ and the 3 bit comparisons of $\mathit{result}$), 4 read only $\pi_1$ (the 3 values of $\mathit{thread}_{\pi_1}$ and $\mathit{mask}_{\pi_1} = 0$), and 1 reads only $\pi_2$ ($\mathit{mask}_{\pi_2} = 0$). The straddling ones are defined over value bits before the product with $\pi_2$.
Eliminating $\pi_2$ then replaces the 4 straddling APs by 6 over $\pi_1$ alone (one per bit of $\mathit{pin}_{\pi_1}$ and of $\mathit{result}_{\pi_1}$) and the AP over $\pi_2$ alone is read inside the product, leaving 10 APs on $\pi_1$. Without compaction, $\mathit{pin}_{\pi_1} \neq \mathit{pin}_{\pi_2}$ would be replaced by 8 APs, one per value of $\mathit{pin}_{\pi_2}$, instead of 3, leaving 15 total.

\paragraph{Automaton reduction.}
As in \autohyper{}, every automaton that a product will read is reduced by \spot{} (\texttt{autfilt --small}).
As this transformation is not idempotent, \cite{hypersmv} introduced iterated reduction until a reaching fixpoint; \tool{} iterates until a round removes less than a tenth of the states.

\paragraph{Complement.}
When the next block's quantifier changes $\mathit{neg}$, every part of $\B$ is negated/complemented and the connective flips.
Where \autohyper{} complements with \spot{}, \tool{} uses \kofola{} by default, which complements modularly~\cite{AlexajHHLLM26}: it partitions the automaton's accepting SCCs by type and complements each partition separately.
\tool{} receives these complements as separate automata, each of which becomes a part of $\overline{\B}$.

Under a union, the input parts are complemented together, as one automaton: their union only adds their sizes, and complementing it as a whole lets \kofola{}'s pruning act across the parts, which yields smaller complements.
Under an intersection, each input part is complemented on its own: complementing them together would first require their product, which multiplies their sizes, before a construction that is exponential in its input.

\paragraph{Product.}
As in \autohyper{}, each block is eliminated by the product of $\B$ with the systems of that block: a product state pairs a state of $\B$ with a state of each eliminated system, and $\B$'s APs are evaluated as in \Cref{sec:background}; a defined AP is evaluated through its definition, whose inputs over the eliminated traces are substituted while those over the remaining traces are kept, so that its residual is the definition over residual inputs.
While building the product, \tool{} also prunes the pairs that lie on no infinite run.

Under a union, each part of $\B$ is taken in product with the systems of the traces of the block it reads.
Under an intersection, the parts that read the block are taken as one product, which reads a tuple of their states.

The product is also where symbolic guards are resolved, and an ordinary NBA over the remaining traces' APs is produced.
A free variable only narrows the letters of the product's edges, as on a system's edges.
A frozen variable must keep its value along the run, which no label can express, and is encoded in one of two ways: enumerated, one product state per value, or symbolically, one product state per set of values still consistent with the run, kept as a decision diagram, much as the set-based reduction of~\cite{BauchHB16}.
When the formula forces the variable equal to a term over other traces, its set is just the value those traces fix, and one symbolic state replaces one state per value; otherwise the system's guards may split its values into many distinct sets, up to one per subset, where enumeration stops at one per value.
\tool{} therefore keeps symbolic exactly the frozen variables the formula determines, and enumerates the others.

In our example, the product of $\B$ with $\pi_2$ pairs states of the conclusion's automaton with dynamic states of $\pi_2$, plus a value of $\mathit{pin}_{\pi_2}$ which the formula does not determine and is therefore enumerated; $\mathit{thread}_{\pi_2}$ is projected away on every edge. Of the $3 \times 305 \times 8$ such triples on \cnilte{3}, 979 are reachable (and 2802 out of $3 \times 606 \times 8$ on \cniany{3}).
Without symbolic edges and formula decomposition, the product pairs the 7 states of the whole body's automaton with the explicit states of $\pi_2$ (745 on \cnilte{3} and 2247 on \cniany{3}), reaching 3705 and 11187 states.

\paragraph{Product quotient.}
\autohyper{} and~\cite{hypersmv} quotient only the systems, once at input; \tool{} also quotients the product built at every iteration, using the same generalised quotient as for the systems.
The two quotients differ in where their information sits.
A system has no acceptance, since \tool{} moves the models' specifications and fairness constraints into the formula, and what the formula reads sits only in states, so quotient classes start from APs alone.
A product inherits acceptance from $\B$, and the residual APs that the formula still reads sits on edge labels, so quotient classes start from acceptance alone.

In our example, the quotient takes the product from 979 to 419 states on \cnilte{3}, and from 2802 to 1069 on \cniany{3}.
Here the product reads only three things of $\pi_2$: the enumerated ${\mathit{pin}}_{\pi_2}$, fixed in the initial state, and both ${\mathit{mask}}_{\pi_2} = 0$ and the bits of ${\mathit{result}}_{\pi_2}$ when $\pi_2$ halts.
Product states that differ only in $\pi_2$'s internal state thus stay together as long as their successors do; and once $\pi_2$ has halted and the eventuality is satisfied, $\pi_2$ no longer tells states apart.

\subsection{Final Check}\label{sec:final}

When there is one block left, the inclusion $\Lang(\A)\subseteq\Lang(\B)$ of $\A$ (that block's self-composition) in $\B$ (the intermediate formula automaton) is done via an external inclusion checker.\footnote{This step is optional, as the loop could run until there are no blocks left and decide with an emptiness check. \autohyper{} and \tool{} also only perform the inclusion check when the input formula has at least one quantifier alternation, otherwise there is no complementation and the emptiness check is often cheaper.}
Nonetheless, \autohyper{} always defaulted to complementation mode when witnesses were to be computed; \cite{hypersmv} lifted the inclusion checkers that natively support counterexamples (\rabit{}, \forklift{} and \roll{}) to map them back to witnesses over the original models.
\tool{} adds novel support for \kofola, also as an inclusion checker.

\paragraph{Inclusion check.}
An inclusion checker receives two automata and explores the product of the first with the complement of the second only as far as its search needs. \autohyper{} and~\cite{hypersmv} hand it $\A$, built explicitly beforehand, whose size is the product of the systems' sizes; the complement of $\B$ will already be computed inside the inclusion checker, often on the fly depending on the particular checking algorithm.
If the outermost quantifier is an $\exists$, the formula is negated first, as the inclusion checker universally quantifies over models.

\tool{} generalizes the inclusion procedure on both sides.
On the left, the automata for separate $\A_{\pi_1},\cdots,\A_{\pi_k}$ are passed directly to \kofola{}, which can now multiply them on the fly as far as its search reaches.
On the right, $\B$ already arrives as separate parts, and behavior depends again on the connective.
Under a union, the parts are unified into one automaton before being passed to \kofola{}. Note that we could move parts of $\B$ to the left (for a part $B$, $\Lang(\A) \subseteq \Lang(B) \cup \Lang(\B) \Leftrightarrow \Lang(\A) \cap \Lang(\overline{B}) \subseteq \Lang(\B)$), but that would require computing their complements beforehand; the exception is a part that still stands for its formula, in which case we move it freely to the left, computing only its syntactic negation.
Under an intersection, the inclusion distributes over it (for a part $B$, $\Lang(\A) \subseteq \Lang(B) \cap \Lang(\B) \Leftrightarrow \Lang(\A) \subseteq \Lang(B) \wedge \Lang(\A) \subseteq \Lang(\B)$) and is checked separately.

This is the only step that hands systems directly to a tool, so each system must become an NBA.
Per trace, the system's automaton keeps its free variables inside its labels.
A frozen variable, which no label can keep constant, is converted into one automaton with a self-looping state per value, so that a run fixes the value once.
The defined APs still in $\B$ come with their definitions. These definitions and the constraints on the frozen variables are conditions on each letter alone. Together they form one more automaton, with one state and a self-loop labelled by their conjunction.
Each premise moved from the formula is translated into an NBA of its own. Those over one trace are multiplied into that trace's system automaton, so that each trace reaches \kofola{} as a single automaton.
$\A$ is then the synchronous product of all these automata, which is passed to \kofola{} as its independent parts (and computed first for the other backends).

Symmetry reduction~\cite{symmetry} exploits renamings of the model's variables that leave model and property unchanged, under which states behave alike; \cite{hypersmv} applied it to the outermost block, before generating the \nusmv{} models. Our frozen variables make it immediate: \tool{} detects the symmetries of the outermost block, which permute its configurations (the valuations of the frozen variables, enumerated by the one-state automaton above), and restricts that automaton to configurations that still cover every class.

In our example, the left operand reaches \kofola{} as four automata: one per bit of ${\mathit{pin}}_{\pi_1}$, and the system of $\pi_1$ multiplied with the three negated fairness conjuncts of the premise, which read $\pi_1$ alone.
The system's automaton has the states of the quotiented system ($305$ on \cnilte{3} and $606$ on \cniany{3}), plus, on \cnilte{3}, 7 initial states that satisfy \texttt{INIT pin > mask} on the first step, since that clause relates the frozen \texttt{pin} to the initial \texttt{mask}.
Multiplied with the NBAs of the three premises, of two states each, it becomes one automaton of 1456 reachable states on \cnilte{3} and 2907 on \cniany{3}.
Their synchronous product, which the other backends receive, has 4559 and 12941 reachable states.

\paragraph{The \kofola{} backend.} The complement at every alternation and the final inclusion check are delegated to \kofola~\cite{AlexajHHLLM26}. Our extension to \kofola{} is minimal: it changes mostly command-line inputs and outputs, and adds no new algorithm.

For inclusion checking, \kofola{} is lazier than for complementation: its search explores the product of the left operand with the complement of the right operand, both built on the fly.
In particular, only letters used in the left operand are followed, and the search stops at the first counterexample.
We allow the left operand to also arrive as separate automata, which then enter the same on-the-fly product.
Original \kofola{} reports only the verdict; we also adapt it to return, on a failed inclusion, a counterexample word.

For complementation, \kofola{} complements each partition of the automaton's SCCs by its own construction, and the complement is the intersection of the partitions' complements; it returns that intersection as one automaton, reduced by \spot's default post-processing. We stop earlier and return each partition's complement as an automaton of its own.

Both modes prune states via a simulation relation on the automaton being complemented. Simulation is computed by one of two standard methods: refining signatures of the states, which are formulas over the alphabet; or refining pairs of states, whose cost follows the number of pairs instead.
\kofola{} takes the relation from \spot's signature-based method, the only one \spot{} exports; over the many APs of our automata, its cost alone can become a bottleneck. We instead compute the relation with the pairwise method, importing what \spot{} implements internally to reduce automata.

\paragraph{Verdict and witness.}
Every inclusion checker that produces a counterexample returns a word that $\A$ accepts and $\B$ rejects.
\cite{hypersmv} recovered the traces from such a word by finding a run of the explicit product $\A$ that reads it; \tool{} instead follows the word through the separate automata of $\A$, so that their product is only built along the concrete word.
Among the resulting runs, the run of each trace's system is then lifted through every quotient to the model's variables, by reachability in the original system as in~\cite{hypersmv}.
Each frozen variable takes its value from the run of its own automaton, and the free variables' values at each step are searched to agree with the step's guard and the word's letter.

In our example, \kofola's counterexample for \cniany{3} is a word over the 10 APs left on $\pi_1$ after eliminating $\pi_2$.
Following it through the four left-side automata yields the lasso of \Cref{fig:example}: ${\mathit{pin}}_{\pi_1} = 001$ is fixed once by the automata of its bits and ${\mathit{thread}}_{\pi_1}$ is chosen at every step from the system's automaton.

\section{Evaluation}\label{sec:eval}

We ask the following questions to assess the feasibility of \tool{}:
\begin{itemize}
\item[\textbf{Q1}] How much does it improve over its ancestor explicit-state model checkers?
\item[\textbf{Q2}] How does it compare with state-of-the-art bounded model checkers?
\item[\textbf{Q3}] Where does it stop and why?
\end{itemize}

\paragraph{Benchmarks.}
For our evaluation, we reuse three published suites from the state of the art, totaling \nrows{} benchmarks.
The first suite (\nrowsHyperQB{} rows) is taken from the most recent \hyperqb{} paper~\cite{hyperqb2}: imperative \nusmv{} models of miscellaneous inspiration.
The second is the ``high'' suite (\nrowsHyperSMV{} rows) of the \hypersmv{} paper~\cite{hypersmv}. Its models were originally written in Alloy and translated automatically into \nusmv{} before being checked by \hypersmv{}, so all of them are declarative \nusmv{} models; some are new, and some redesign earlier \hyperqb{} benchmarks.
The third suite (\nrowsHyperLasso{} rows), from \hyperlasso{}~\cite{hyperlasso}, also consists of declarative models. It focuses on liveness hyperproperties; some are new, and some extend earlier benchmarks with liveness.
Almost every row has a single quantifier alternation (only two \hyperqb{} rows have more), so the complementation at an inner quantifier alternation will be rarely exercised by these suites (as already happened for our running example).

\paragraph{Setup.}
We ran three explicit-state tools (\autohyper~\cite{BeutnerF23}, \hypersmv~\cite{hypersmv} with its customised \autohyper{} backend, and \tool{}) and three symbolic ones (\hyperqb~\cite{hyperqb2} with its SMT backend, \hypersmv~\cite{hypersmv} with its QBF backend, and \hyperlasso~\cite{hyperlasso}).
We considered the latest version in the GitHub repository given in each tool's paper.
All experiments ran on a MacBook Air with an Apple M5 chip and 16 GB of memory, with a wall-clock timeout of \timeoutsecs\,s per run.
The two \autohyper{} variants use, on each row, the automata backend its suite prescribes; \tool{} uses \kofola{} on every row.
The symbolic tools were invoked with bounded semantics and unrolling bound as prescribed in the suites.
Every tool was asked for a witness when one is expected.

\paragraph{Results.}
\Cref{tab:results} presents the results of every tool on all benchmarks from the three suites.
Each row is one benchmark, grouped by suite.
Column \textbf{Q*} gives the quantifier prefix;
\textbf{Magnitude} the decimal exponent of the number of reachable states of the product of the models, computed by \nuxmv{};
\textbf{Res} the expected verdict (✓ holds, ✗ violated), as given by the suite.
Each tool's column gives the time in seconds or a labeled result.
All the errors reported for \autohyper{} and \hyperqb{} are because they do not support declarative \nusmv models.
The symbolic tools are bounded, and we only report their conclusive results.
\hyperqb{} and \hypersmv{} are only conclusive when the bounded semantics is opposite to the verdict, or when the system halts before the maximum unrolling bound, for any verdict.
\hyperlasso{} is only conclusive for up to one quantifier alternation, when it produces a witness.

\paragraph{Q1: \tool{} vs explicit tools.}
\tool{} decides all rows that one of its ancestors decides except one, and 22 rows that neither decides.
Many of the 22 are the larger scopes of families, where ancestors only decide the smallest instances and the growth is tackled by the novel \tool{}'s techniques.

Three such growth sources recur.
The first is inputs. The declarative suites encode a model's configuration and environment choices as frozen and free variables: while an explicit-state tool multiplies every state by their valuations, symbolic edges keep them off the states. In the CNI family of the \hyperlasso{} suite, when the initial mask is arbitrary neither ancestor decides beyond the second scope, while \tool{}'s time grows only about $10\times$ from smallest to largest scope.
The second is the left side of the final check, the outermost block. Its systems and the automata of their frozen variables reach \kofola{} as separate parts, which multiplies them only as far as its search needs.
In the CMS family of the \hyperlasso{} suite, whose outermost block has two universal traces, both ancestors decide only the smallest scope, and \tool{} also decides the next one, about nine orders of magnitude larger.
Unlike the CMS models of the \hypersmv{} suite, whose outer blocks where already self-composed with their symmetries trimmed, similar models from the \hyperlasso{} suite have much larger magnitudes; \tool{} detects one symmetry itself, up to halving the configurations the search starts from.
The third is the right side of the final check, the innermost automaton. Decomposition splits it into parts checked separately, the product quotient shrinks each part, and \kofola{} complements them only as far as its search needs. In the Planning family of the \hyperlasso{} suite, whose formulas have as many innermost quantifiers as robots, decomposition yields one part per robot: the \hypersmv{} version of \autohyper{} becomes almost $100\times$ slower from one universal trace to three and fails with four robots, while \tool{}'s time stays the same.

In the CMS $\text{any}_\textsc{GNI}$ family of the \hypersmv{} suite, which includes the $3\!\times\!2$ row that \tool{} loses, \hypersmv{} splits the initial configurations of the model and checks one at a time, which allows it to stop at the first configuration that yields a counterexample; it loses however the other $3\!\times\!2$ GNI rows, of slightly smaller magnitude, but where all configurations need to be explored, and for which \tool{} does complete.

The SNARK family of the \hypersmv{} suite shows the widest gap to the earlier explicit-state tools, that time out, while \tool{} decides each within seconds. \tool{} decides even with the inclusion checkers supported by \hypersmv{}, and \kofola{} makes it about an order of magnitude faster.

Although \kofola{} contributes substantially overall, the other techniques remain key for several benchmarks.
For inputs, switching off symbolic edges makes the largest CNI row about $3\times$ slower, the largest Self-stabilisation row about $20\times$ slower, and the largest row of the first Controller family run out of memory.
For the outermost block, the largest \hyperlasso{} and \hypersmv{} CMS rows that \tool{} decides run out of memory when their left operand is composed up front instead of being handed to \kofola{} as separate parts.
For the innermost automaton, the largest Planning row times out without decomposition, already while translating the full body, compared to under a second with decomposition.

\paragraph{Q2: \tool{} vs symbolic tools.}
The bounded tools and \tool{} answer different questions, and~\Cref{tab:results} shows this more than any difference in speed. The bounded tools are also often dual to each other: \hyperlasso{} is the only one conclusive when the witness is an infinite trace, and \hyperqb{} and \hypersmv{} are the only ones that can prove that no witness exists, but only for simple properties or terminating systems.
When \hyperqb{} and \hypersmv{} decide, they are typically much faster than \hyperlasso{}, as they don't pay the added cost of encoding loops.
In either case, the cost of all symbolic tools follows the unrolling depth, which caps the length of the witness and determines the size of its encoding. \tool{}'s cost is the opposite: it follows the reachable state space of the constructed automata, whatever the property or the verdict.

Of the 37 rows that \tool{} decides and no bounded tool does, 34 need a proof that no witness exists, which a bounded tool could only give by unrolling up to a completeness threshold far beyond its reach. The other 3, from the \hyperqb{} suite, need either a long unrolling or multiple quantifier alternations. In the Robot short family, the unrolling bound of \hyperlasso{} grows with the scope, $6\times$ from the smallest to the largest, and it times out on the two largest scopes, while \tool{}'s time grows only about $5\times$ across the family.\footnote{Remark that, in both Robot families of the \hyperqb{} suite, the number of states does not grow proportionally to the scope, with magnitudes between 5 and 7.} Mapping synthesis has two quantifier alternations.

Conversely, the 4 rows that a bounded tool decides and \tool{} does not, from the CMS, Isolation and Controller \hyperlasso{} families, all have short witnesses on very large models, of magnitude 12 to 30. For the same reason, on the violated CMS rows that both tool categories decide, a symbolic tool is up to two orders of magnitude faster than \tool{}. The two approaches are therefore complementary rather than competing: together, \tool{} and \hyperlasso{} decide every row that any tool decides.

\paragraph{Q3: \tool{} limitations.}
\tool{} leaves \the\numexpr\nrows-\solvedOmegaHyperHyperQB-\solvedOmegaHyperHyperLasso-\solvedOmegaHyperHyperSMV\relax{} rows undecided, which mark the size limits of the approach: 3 of them are satisfied universal properties of magnitude 26 to 30 that no tool decides; the other 4 have the short witnesses discussed above, but with models of significant magnitudes between 12 and 30.

The first size limit is in the dynamic system. \tool{} handles symbolically the variables that do not evolve, but whatever evolves must still be enumerated. On the CMS $3\!\times\!3$ rows of the \hyperlasso{} suite, of magnitude 26 to 28, the dynamic system alone is too large, and the translation into an explicit system runs out of memory or time before any check begins. No reduction applied after the translation can help here, and going forward would take some compositional exploration of the models.

The second size limit is in the frozen variables. Symbolic edges keep them off the states, but the final check enumerates their valuations, and when not constrained severely by the formula their number multiplies the cost. This is what stops the largest Controller and Isolation rows, of magnitude 16 and 30. Symmetry breaking shrinks the number of configurations, but not its exponential growth; splitting models by their configurations, and checking them one at a time, as \hypersmv{} does, only pays off when a witness is found in one of the early splits.
One way forward would be to keep the frozen variables symbolic, which explicit-state inclusion checkers do not support today.

\begin{table}[p]
\centering\tiny\setlength{\aboverulesep}{0pt}\setlength{\belowrulesep}{0pt}\def\rowstrut{\vphantom{$\forall^{2}\exists$}\vphantom{A$_{\text{\textsc{x}}}10^{2}$}\vphantom{✓✗}}\setlength{\tabcolsep}{0.8pt}\renewcommand{\arraystretch}{1}\newcolumntype{T}{w{c}{\toolcolw}}
\caption{Execution results for the 3 benchmark suites under the 6 tools. For each tool column: time in seconds, ⏱~timeout, 🔋~memory-out, ❗~error, \na{}~inconclusive Background color per row: \protect\swatch{fastAll}~fastest overall; \protect\swatch{fastGroup}~fastest of its tool category (explicit or symbolic). }\label{tab:results}

\medskip
\begin{tabular}[t]{!{\vrule\hspace{1pt}}c!{\hspace{1pt}\vrule\hspace{1pt}}l;{.4pt/.8pt}c;{.4pt/.8pt}c;{.4pt/.8pt}c!{\hspace{1pt}\vrule\hspace{1pt}}T;{.4pt/.8pt}T;{.4pt/.8pt}T!{\hspace{1pt}\vrule\hspace{1pt}}T;{.4pt/.8pt}T;{.4pt/.8pt}T!{\hspace{1pt}\vrule}}
\toprule
\multicolumn{1}{!{\vrule\hspace{1pt}}c!{\hspace{1pt}\vrule\hspace{1pt}}}{} &  & & & & \multicolumn{3}{c!{\hspace{1pt}\vrule\hspace{1pt}}}{explicit} & \multicolumn{3}{c!{\hspace{1pt}\vrule}}{symbolic}\\
\rotatebox{90}{\textbf{Suite}} & \textbf{Benchmark} & \textbf{Q*} & \rotatebox{90}{\textbf{Magnitude}} & \rotatebox{90}{\textbf{Res}} & \rotatebox{90}{\autohyper} & \rotatebox{90}{\hypersmv} & \rotatebox{90}{\tool} & \rotatebox{90}{\hyperqb} & \rotatebox{90}{\hypersmv} & \rotatebox{90}{\hyperlasso} \\
\midrule
\multirow{47}{*}{\rotatebox[origin=c]{90}{\textbf{HyperQB}}} & \rowstrut{}\bakery{3} & $\forall\exists$ & 4 & ✗
& \fa{0.2} & 0.3 & 0.3 & \na & \na & \fg{0.4} \\
\cdashline{2-11}[.4pt/.8pt]
 & \rowstrut{}\bakery{7} & $\forall\exists$ & 7 & ✗
 & 2.4 & 5.8 & \fa{0.4} & \na & \na & \fg{1.2} \\
\cdashline{2-11}[.4pt/.8pt]
 & \rowstrut{}\bakery{9} & $\forall\exists$ & 6 & ✗
 & 4.1 & 23 & \fa{0.5} & \na & \na & \fg{23} \\
\cdashline{2-11}[.4pt/.8pt]
 & \rowstrut{}\bakery{11} & $\forall\exists$ & 7 & ✗
 & 24 & 377 & \fa{0.8} & \na & \na & \fg{145} \\
\cdashline{2-11}[.4pt/.8pt]
 & \rowstrut{}\snark{} & $\forall\exists$ & 6 & ✗
 & 3.2 & 1.5 & \fa{0.6} & \na & \na & \fg{8.2} \\
\cdashline{2-11}[.4pt/.8pt]
 & \rowstrut{}\nicorrect{} & $\forall\exists$ & 4 & ✓
 & \fa{0.2} & \fa{0.2} & 0.3 & \fg{0.4} & 0.7 & \na \\
\cdashline{2-11}[.4pt/.8pt]
 & \rowstrut{}\niincorrect{} & $\forall\exists$ & 2 & ✗
 & \fg{0.2} & \fg{0.2} & \fg{0.2} & \fa{0.1} & 0.6 & 29 \\
\cdashline{2-11}[.4pt/.8pt]
 & \rowstrut{}\nrpcorrect{} & $\exists\forall$ & 4 & ✓
 & \fa{0.2} & \fa{0.2} & 0.3 & \na & \na & \fg{0.5} \\
\cdashline{2-11}[.4pt/.8pt]
 & \rowstrut{}\nrpincorrect{} & $\exists\forall$ & 3 & ✗
 & 0.3 & \fg{0.2} & 0.3 & \fa{0.0} & 0.2 & \na \\
\cdashline{2-11}[.4pt/.8pt]
 & \rowstrut{}\robotrobust[10] & $\exists\forall$ & 5 & ✓
 & ⏱ & 2.1 & \fg{0.3} & \fa{0.2} & 0.4 & 1.8 \\
\cdashline{2-11}[.4pt/.8pt]
 & \rowstrut{}\robotrobust[20] & $\exists\forall$ & 6 & ✓
 & 🔋 & 6.5 & \fa{0.4} & \fg{1.6} & 1.8 & 15 \\
\cdashline{2-11}[.4pt/.8pt]
 & \rowstrut{}\robotrobust[40] & $\exists\forall$ & 7 & ✓
 & 🔋 & 19 & \fa{0.9} & 4.6 & \fg{3.7} & 42 \\
\cdashline{2-11}[.4pt/.8pt]
 & \rowstrut{}\robotrobust[60] & $\exists\forall$ & 7 & ✓
 & 318 & 12 & \fa{1.1} & 366 & \fg{16} & 201 \\
\cdashline{2-11}[.4pt/.8pt]
 & \rowstrut{}\robotshort[10] & $\exists\forall$ & 6 & ✓
 & 2.2 & 1.8 & \fa{0.4} & \na & \na & \fg{4.1} \\
\cdashline{2-11}[.4pt/.8pt]
 & \rowstrut{}\robotshort[20] & $\exists\forall$ & 6 & ✓
 & 2.0 & 0.7 & \fa{0.5} & \na & \na & \fg{42} \\
\cdashline{2-11}[.4pt/.8pt]
 & \rowstrut{}\robotshort[40] & $\exists\forall$ & 7 & ✓
 & 51 & 🔋 & \fa{1.8} & \na & \na & ⏱ \\
\cdashline{2-11}[.4pt/.8pt]
 & \rowstrut{}\robotshort[60] & $\exists\forall$ & 6 & ✓
 & 15 & 🔋 & \fa{2.1} & \na & \na & ⏱ \\
\cdashline{2-11}[.4pt/.8pt]
 & \rowstrut{}\mutant{} & $\exists\forall$ & 3 & ✓
 & \fg{0.2} & 0.3 & 0.3 & \fa{0.0} & 0.2 & 0.2 \\
\cdashline{2-11}[.4pt/.8pt]
 & \rowstrut{}\coterm{} & $\forall^{2}$ & 3 & ✓
 & \fg{0.2} & 0.4 & \fg{0.2} & \fa{0.0} & 0.7 & \na \\
\cdashline{2-11}[.4pt/.8pt]
 & \rowstrut{}\deniability{} & $\forall\exists^{2}$ & 10 & ✓
 & 8.1 & 3.8 & \fa{0.7} & 32 & \fg{2.0} & \na \\
\cdashline{2-11}[.4pt/.8pt]
 & \rowstrut{}\bufscod{} & $\forall^{2}$ & 5 & ✗
 & \fg{0.2} & \fg{0.2} & \fg{0.2} & \fa{0.0} & 0.2 & 0.2 \\
\cdashline{2-11}[.4pt/.8pt]
 & \rowstrut{}\bufsiod{} & $\forall^{2}$ & 5 & ✓
 & 0.4 & \fg{0.3} & \fg{0.3} & \fa{0.0} & 0.3 & \na \\
\cdashline{2-11}[.4pt/.8pt]
 & \rowstrut{}\bufsigmni{} & $\forall\exists$ & 5 & ✓
 & 1.5 & \fg{0.3} & \fg{0.3} & \fa{0.1} & 0.3 & \na \\
\cdashline{2-11}[.4pt/.8pt]
 & \rowstrut{}\bufucod{} & $\forall^{2}$ & 6 & ✗
 & 0.3 & \fg{0.2} & 0.3 & \fa{0.0} & 0.2 & 0.1 \\
\cdashline{2-11}[.4pt/.8pt]
 & \rowstrut{}\nodettini{} & $\forall\exists$ & 6 & ✓
 & 0.5 & \fg{0.3} & \fg{0.3} & \fa{0.0} & 0.2 & \na \\
\cdashline{2-11}[.4pt/.8pt]
 & \rowstrut{}\nodettsni{} & $\forall\exists$ & 6 & ✓
 & 0.5 & \fg{0.3} & \fg{0.3} & \fa{0.0} & 0.2 & \na \\
\cdashline{2-11}[.4pt/.8pt]
 & \rowstrut{}\ksafety{} & $\forall^{2}$ & 5 & ✓
 & 0.8 & 0.4 & \fa{0.3} & \fg{1.2} & \fg{1.2} & \na \\
\cdashline{2-11}[.4pt/.8pt]
 & \rowstrut{}\mapsynth{1} & $\exists\forall^{2}\exists^{2}$ & 5 & ✓
 & \fg{0.2} & \fg{0.2} & 0.3 & \fa{0.0} & 0.5 & \na \\
\cdashline{2-11}[.4pt/.8pt]
 & \rowstrut{}\mapsynth{2} & $\exists\forall^{2}\exists^{2}$ & 10 & ✓
 & 🔋 & 49 & \fa{0.8} & \na & \na & \na \\
\cdashline{2-11}[.4pt/.8pt]
 & \rowstrut{}\teamltl{1} & $\exists^{2}\forall$ & 5 & ✗
 & \fa{0.2} & 2.4 & 0.3 & 0.5 & \fg{0.3} & \na \\
\cdashline{2-11}[.4pt/.8pt]
 & \rowstrut{}\teamltl{2} & $\exists^{2}\forall$ & 6 & ✗
 & \fa{0.3} & 49 & \fa{0.3} & 7.6 & \fg{0.4} & \na \\
\cdashline{2-11}[.4pt/.8pt]
 & \rowstrut{}\ndet{1} & $\forall\exists$ & 2 & ✗
 & \fg{0.2} & \fg{0.2} & 0.3 & \fa{0.0} & 0.2 & 0.1 \\
\cdashline{2-11}[.4pt/.8pt]
 & \rowstrut{}\ndet{2} & $\forall\exists$ & 3 & ✗
 & \fg{0.2} & \fg{0.2} & 0.3 & \fa{0.0} & 0.2 & 0.2 \\
\cdashline{2-11}[.4pt/.8pt]
 & \rowstrut{}\ndet{3} & $\forall\exists$ & 3 & ✗
 & 42 & \fg{0.3} & \fg{0.3} & \fa{0.0} & 0.2 & 0.2 \\
\cdashline{2-11}[.4pt/.8pt]
 & \rowstrut{}\bank{1} & $\forall^{2}\exists$ & 9 & ✗
 & \fg{0.3} & \fg{0.3} & \fg{0.3} & \fa{0.0} & 0.3 & 0.2 \\
\cdashline{2-11}[.4pt/.8pt]
 & \rowstrut{}\bank{2} & $\forall^{2}\exists$ & 7 & ✗
 & \fg{0.2} & 0.3 & 0.3 & \fa{0.0} & 0.3 & 0.2 \\
\cdashline{2-11}[.4pt/.8pt]
 & \rowstrut{}\bank{3} & $\forall^{2}\exists$ & 6 & ✗
 & \fg{0.2} & \fg{0.2} & 0.3 & \fa{0.0} & 0.3 & 0.2 \\
\cdashline{2-11}[.4pt/.8pt]
 & \rowstrut{}\constructor{} & $\forall\exists$ & 4 & ✓
 & \fa{0.2} & \fa{0.2} & 0.3 & \na & \na & \na \\
\cdashline{2-11}[.4pt/.8pt]
 & \rowstrut{}\bidding{1} & $\forall^{2}$ & 2 & ✓
 & \fg{0.2} & \fg{0.2} & \fg{0.2} & \fa{0.0} & 0.2 & \na \\
\cdashline{2-11}[.4pt/.8pt]
 & \rowstrut{}\bidding{2} & $\forall^{2}$ & 3 & ✓
 & \fa{0.2} & \fa{0.2} & \fa{0.2} & \na & \na & \na \\
\cdashline{2-11}[.4pt/.8pt]
 & \rowstrut{}\bidding{3} & $\forall^{2}$ & 6 & ✓
 & 0.4 & \fa{0.2} & 0.3 & \na & \na & \na \\
\cdashline{2-11}[.4pt/.8pt]
 & \rowstrut{}\bidding{4} & $\forall^{2}$ & 4 & ✗
 & \fg{0.2} & \fg{0.2} & \fg{0.2} & \fa{0.0} & 0.2 & 0.1 \\
\cdashline{2-11}[.4pt/.8pt]
 & \rowstrut{}\iqueue{} & $\forall\exists$ & 8 & ✗
 & 🔋 & 16 & \fa{1.0} & \na & \na & \fg{32} \\
\cdashline{2-11}[.4pt/.8pt]
 & \rowstrut{}\keypad{} & $\exists^{2}$ & 5 & ✓
 & 10 & 3.0 & \fg{2.2} & \fa{0.0} & 0.2 & 0.1 \\
\cdashline{2-11}[.4pt/.8pt]
 & \rowstrut{}\simplequeue{} & $\forall^{2}$ & 2 & ✗
 & \fg{0.2} & \fg{0.2} & 0.3 & \fa{0.0} & 0.3 & 0.2 \\
\cdashline{2-11}[.4pt/.8pt]
 & \rowstrut{}\emmaba{} & $\forall\exists$ & 6 & ✗
 & 🔋 & 2.5 & \fg{0.4} & \fa{0.1} & 0.3 & 0.4 \\
\cdashline{2-11}[.4pt/.8pt]
 & \rowstrut{}\lazylist{} & $\forall\exists$ & 7 & ✗
 & 476 & 8.1 & \fg{1.2} & \fa{0.2} & 0.5 & 1.6 \\
\midrule
\multirow{13}{*}{\rotatebox[origin=c]{90}{\textbf{HyperSMV}}} & \rowstrut{}\mutantpk{1} & $\forall\exists$ & 3 & ✗
& ❗ & \fa{0.2} & 0.4 & ❗ & \fa{0.2} & 0.9 \\
\cdashline{2-11}[.4pt/.8pt]
& \rowstrut{}\mutantdk{1} & $\forall\exists$ & 3 & ✗
& ❗ & \fa{0.2} & 0.3 & ❗ & \fa{0.2} & \fa{0.2} \\
\cdashline{2-11}[.4pt/.8pt]
& \rowstrut{}\mutantpk{2} & $\forall\exists$ & 3 & ✗
& ❗ & \fa{0.2} & 0.3 & ❗ & \fa{0.2} & 0.3 \\
\cdashline{2-11}[.4pt/.8pt]
& \rowstrut{}\mutantdk{2} & $\forall\exists$ & 4 & ✓
& ❗ & \fa{0.2} & 0.3 & \na & \na & \na \\
\cdashline{2-11}[.4pt/.8pt]
& \rowstrut{}\mutantpk{3} & $\forall\exists$ & 2 & ✓
& ❗ & \fa{0.2} & 0.3 & \na & \na & \na \\
\cdashline{2-11}[.4pt/.8pt]
& \rowstrut{}\mutantdk{3} & $\forall\exists$ & 3 & ✓
& ❗ & \fa{0.2} & 0.3 & \na & \na & \na \\
\cdashline{2-11}[.4pt/.8pt]
& \rowstrut{}\robottwo{} & $\forall^{2}$ & 5 & ✗
& ❗ & 19 & \fa{1.1} & ❗ & \fg{3.0} & 9.9 \\
\cdashline{2-11}[.4pt/.8pt]
& \rowstrut{}\robotshort{} & $\forall\exists$ & 4 & ✗
& ❗ & 0.6 & \fa{0.5} & \na & \na & \fg{12} \\
\cdashline{2-11}[.4pt/.8pt]
& \rowstrut{}\robotscore{} & $\forall\exists$ & 6 & ✗
& ❗ & 2.1 & \fa{0.8} & \na & \na & \fg{11} \\
\cdashline{2-11}[.4pt/.8pt]
& \rowstrut{}\robotrobust{} & $\forall\exists$ & 5 & ✗
& ❗ & 1.0 & \fa{0.6} & \na & \na & \fg{5.4} \\
\cdashline{2-11}[.4pt/.8pt]
& \rowstrut{}\robotadv{} & $\forall\exists$ & 4 & ✗
& ❗ & \fa{1.1} & 1.2 & \na & \na & \fg{6.1} \\
\cdashline{2-11}[.4pt/.8pt]
& \rowstrut{}\cmsanyni{2}{2} & $\forall$ & 4 & ✗
& ❗ & \fg{0.7} & 0.9 & ❗ & \fa{0.2} & \fa{0.2} \\
\cdashline{2-11}[.4pt/.8pt]
& \rowstrut{}\cmsanyni{3}{2} & $\forall$ & 6 & ✗
& ❗ & 139 & \fg{32} & ❗ & \fa{0.4} & 0.6 \\
\bottomrule
\end{tabular}
\hfill
\begin{tabular}[t]{!{\vrule\hspace{1pt}}c!{\hspace{1pt}\vrule\hspace{1pt}}l;{.4pt/.8pt}c;{.4pt/.8pt}c;{.4pt/.8pt}c!{\hspace{1pt}\vrule\hspace{1pt}}T;{.4pt/.8pt}T;{.4pt/.8pt}T!{\hspace{1pt}\vrule\hspace{1pt}}T;{.4pt/.8pt}T;{.4pt/.8pt}T!{\hspace{1pt}\vrule}}
\toprule
\multicolumn{1}{!{\vrule\hspace{1pt}}c!{\hspace{1pt}\vrule\hspace{1pt}}}{} &  & & & & \multicolumn{3}{c!{\hspace{1pt}\vrule\hspace{1pt}}}{explicit} & \multicolumn{3}{c!{\hspace{1pt}\vrule}}{symbolic}\\
\rotatebox{90}{\textbf{Suite}} & \textbf{Benchmark} & \textbf{Q*} & \rotatebox{90}{\textbf{Magnitude}} & \rotatebox{90}{\textbf{Res}} & \rotatebox{90}{\autohyper} & \rotatebox{90}{\hypersmv} & \rotatebox{90}{\tool} & \rotatebox{90}{\hyperqb} & \rotatebox{90}{\hypersmv} & \rotatebox{90}{\hyperlasso} \\
\midrule
\multirow{12}{*}{\rotatebox[origin=c]{90}{\textbf{HyperSMV}}} & \rowstrut{}\cmsanygni{2}{2} & $\forall\exists$ & 8 & ✗
& ❗ & \fg{1.6} & 4.7 & ❗ & \fa{0.4} & 0.7 \\
\cdashline{2-11}[.4pt/.8pt]
& \rowstrut{}\cmsanygni{3}{2} & $\forall\exists$ & 12 & ✗
& ❗ & \fg{128} & 🔋 & ❗ & \fa{0.7} & 1.5 \\
\cdashline{2-11}[.4pt/.8pt]
& \rowstrut{}\cmsndetni{2}{2} & $\forall$ & 3 & ✗
& ❗ & \fg{0.5} & 0.6 & ❗ & \fa{0.3} & \fa{0.3} \\
\cdashline{2-11}[.4pt/.8pt]
& \rowstrut{}\cmsndetni{3}{2} & $\forall$ & 6 & ✗
& ❗ & 83 & \fg{15} & ❗ & \fa{0.4} & 0.6 \\
\cdashline{2-11}[.4pt/.8pt]
& \rowstrut{}\cmsndetgni{2}{2} & $\forall\exists$ & 7 & ✓
& ❗ & 3.4 & \fa{2.8} & \na & \na & \na \\
\cdashline{2-11}[.4pt/.8pt]
& \rowstrut{}\cmsndetgni{3}{2} & $\forall\exists$ & 11 & ✓
& ❗ & 🔋 & \fa{797} & \na & \na & \na \\
\cdashline{2-11}[.4pt/.8pt]
& \rowstrut{}\cmsmaxni{2}{2} & $\forall$ & 3 & ✓
& ❗ & \fa{0.4} & 0.5 & \na & \na & \na \\
\cdashline{2-11}[.4pt/.8pt]
& \rowstrut{}\cmsmaxni{3}{2} & $\forall$ & 5 & ✓
& ❗ & 39 & \fa{6.9} & \na & \na & \na \\
\cdashline{2-11}[.4pt/.8pt]
& \rowstrut{}\cmsmaxgni{2}{2} & $\forall\exists$ & 7 & ✓
& ❗ & 3.0 & \fa{2.3} & \na & \na & \na \\
\cdashline{2-11}[.4pt/.8pt]
& \rowstrut{}\cmsmaxgni{3}{2} & $\forall\exists$ & 10 & ✓
& ❗ & ⏱ & \fa{254} & \na & \na & \na \\
\cdashline{2-11}[.4pt/.8pt]
& \rowstrut{}\snarkbug{} & $\forall\exists$ & 10 & ✗
& ❗ & ⏱ & \fa{9.6} & ❗ & \fg{19} & 71 \\
\cdashline{2-11}[.4pt/.8pt]
& \rowstrut{}\snarkfix{} & $\forall\exists$ & 10 & ✓
& ❗ & ⏱ & \fa{11} & \na & \na & \na \\
\midrule
\multirow{48}{*}{\rotatebox[origin=c]{90}{\textbf{HyperLasso}}} & \rowstrut{}\cmsanygni{1}{1} & $\forall^{2}\exists$ & 6 & ✓
& 23 & 0.5 & \fa{0.3} & \na & \na & \na \\
\cdashline{2-11}[.4pt/.8pt]
& \rowstrut{}\cmsanygni{2}{2} & $\forall^{2}\exists$ & 15 & ✗
& ⏱ & ⏱ & \fg{729} & \na & \na & \fa{0.5} \\
\cdashline{2-11}[.4pt/.8pt]
& \rowstrut{}\cmsanygni{3}{3} & $\forall^{2}\exists$ & 28 & ✗
& 🔋 & ⏱ & ⏱ & \na & \na & \fa{1.7} \\
\cdashline{2-11}[.4pt/.8pt]
& \rowstrut{}\cmsndetgni{1}{1} & $\forall^{2}\exists$ & 6 & ✓
& 27 & 0.5 & \fa{0.4} & \na & \na & \na \\
\cdashline{2-11}[.4pt/.8pt]
& \rowstrut{}\cmsndetgni{2}{2} & $\forall^{2}\exists$ & 14 & ✓
& ⏱ & ⏱ & \fa{84} & \na & \na & \na \\
\cdashline{2-11}[.4pt/.8pt]
& \rowstrut{}\cmsndetgni{3}{3} & $\forall^{2}\exists$ & 26 & ✓
& ⏱ & ⏱ & 🔋 & \na & \na & \na \\
\cdashline{2-11}[.4pt/.8pt]
& \rowstrut{}\cmsmaxgni{1}{1} & $\forall^{2}\exists$ & 6 & ✓
& 24 & \fa{0.5} & 0.9 & \na & \na & \na \\
\cdashline{2-11}[.4pt/.8pt]
& \rowstrut{}\cmsmaxgni{2}{2} & $\forall^{2}\exists$ & 14 & ✓
& ⏱ & ⏱ & \fa{73} & \na & \na & \na \\
\cdashline{2-11}[.4pt/.8pt]
& \rowstrut{}\cmsmaxgni{3}{3} & $\forall^{2}\exists$ & 26 & ✓
& ⏱ & ⏱ & 🔋 & \na & \na & \na \\
\cdashline{2-11}[.4pt/.8pt]
& \rowstrut{}\isolrcser{2}{1} & $\forall\exists$ & 4 & ✓
& \fa{0.2} & \fa{0.2} & 0.5 & \na & \na & \na \\
\cdashline{2-11}[.4pt/.8pt]
& \rowstrut{}\isolrcser{3}{2} & $\forall\exists$ & 13 & ✗
& 🔋 & 🔋 & \fg{8.0} & \na & \na & \fa{0.4} \\
\cdashline{2-11}[.4pt/.8pt]
& \rowstrut{}\isolrcser{4}{3} & $\forall\exists$ & 30 & ✗
& 🔋 & 🔋 & 🔋 & \na & \na & \fa{1.2} \\
\cdashline{2-11}[.4pt/.8pt]
& \rowstrut{}\isolserrc{2}{1} & $\forall\exists$ & 4 & ✓
& \fa{0.2} & \fa{0.2} & 0.5 & \na & \na & \na \\
\cdashline{2-11}[.4pt/.8pt]
& \rowstrut{}\isolserrc{3}{2} & $\forall\exists$ & 13 & ✓
& 🔋 & 🔋 & \fa{10} & \na & \na & \na \\
\cdashline{2-11}[.4pt/.8pt]
& \rowstrut{}\isolserrc{4}{3} & $\forall\exists$ & 30 & ✓
& 🔋 & 🔋 & 🔋 & \na & \na & \na \\
\cdashline{2-11}[.4pt/.8pt]
& \rowstrut{}\cniany{1} & $\forall\exists$ & 5 & ✗
& \fa{0.3} & 0.8 & 0.4 & \na & \na & \fg{0.5} \\
\cdashline{2-11}[.4pt/.8pt]
& \rowstrut{}\cniany{2} & $\forall\exists$ & 6 & ✗
& 12 & ⏱ & \fa{0.4} & \na & \na & \fg{1.5} \\
\cdashline{2-11}[.4pt/.8pt]
& \rowstrut{}\cniany{3} & $\forall\exists$ & 8 & ✗
& ⏱ & 🔋 & \fa{0.7} & \na & \na & \fg{1.8} \\
\cdashline{2-11}[.4pt/.8pt]
& \rowstrut{}\cniany{4} & $\forall\exists$ & 9 & ✗
& ⏱ & ⏱ & \fg{3.1} & \na & \na & \fa{2.2} \\
\cdashline{2-11}[.4pt/.8pt]
& \rowstrut{}\cnilte{1} & $\forall\exists$ & 1 & ✗
& ❗ & 🔋 & \fa{0.3} & \na & \na & \fa{0.3} \\
\cdashline{2-11}[.4pt/.8pt]
& \rowstrut{}\cnilte{2} & $\forall\exists$ & 5 & ✗
& ❗ & ⏱ & \fa{0.3} & \na & \na & \fg{2.5} \\
\cdashline{2-11}[.4pt/.8pt]
& \rowstrut{}\cnilte{3} & $\forall\exists$ & 7 & ✓
& ❗ & 1.1 & \fa{0.5} & \na & \na & \na \\
\cdashline{2-11}[.4pt/.8pt]
& \rowstrut{}\cnilte{4} & $\forall\exists$ & 9 & ✓
& ❗ & 12 & \fa{2.4} & \na & \na & \na \\
\cdashline{2-11}[.4pt/.8pt]
& \rowstrut{}\selfstablf{3} & $\forall\exists$ & 3 & ✓
& ❗ & \fa{0.2} & 0.3 & \na & \na & \na \\
\cdashline{2-11}[.4pt/.8pt]
& \rowstrut{}\selfstablf{5} & $\forall\exists$ & 5 & ✓
& ❗ & 🔋 & \fa{0.4} & \na & \na & \na \\
\cdashline{2-11}[.4pt/.8pt]
& \rowstrut{}\selfstablf{7} & $\forall\exists$ & 6 & ✓
& ❗ & ⏱ & \fa{5.8} & \na & \na & \na \\
\cdashline{2-11}[.4pt/.8pt]
& \rowstrut{}\selfstabgf{3} & $\forall\exists$ & 3 & ✗
& ❗ & \fa{0.3} & \fa{0.3} & \na & \na & \fg{0.4} \\
\cdashline{2-11}[.4pt/.8pt]
& \rowstrut{}\selfstabgf{5} & $\forall\exists$ & 5 & ✗
& ❗ & 5.5 & \fg{0.4} & \na & \na & \fa{0.2} \\
\cdashline{2-11}[.4pt/.8pt]
& \rowstrut{}\selfstabgf{7} & $\forall\exists$ & 6 & ✗
& ❗ & 🔋 & \fg{0.5} & \na & \na & \fa{0.2} \\
\cdashline{2-11}[.4pt/.8pt]
& \rowstrut{}\robustcrit{1} & $\forall\exists$ & 2 & ✓
& ❗ & \fa{0.2} & 0.3 & \na & \na & \na \\
\cdashline{2-11}[.4pt/.8pt]
& \rowstrut{}\robustcrit{2} & $\forall\exists$ & 4 & ✗
& ❗ & \fa{0.3} & \fa{0.3} & \na & \na & \fg{0.6} \\
\cdashline{2-11}[.4pt/.8pt]
& \rowstrut{}\robustcrit{3} & $\forall\exists$ & 7 & ✗
& ❗ & 1.7 & \fg{0.5} & \na & \na & \fa{0.3} \\
\cdashline{2-11}[.4pt/.8pt]
& \rowstrut{}\robustncrit{1} & $\forall\exists$ & 2 & ✓
& ❗ & \fa{0.2} & 0.3 & \na & \na & \na \\
\cdashline{2-11}[.4pt/.8pt]
& \rowstrut{}\robustncrit{2} & $\forall\exists$ & 4 & ✓
& ❗ & 4.9 & \fa{0.3} & \na & \na & \na \\
\cdashline{2-11}[.4pt/.8pt]
& \rowstrut{}\robustncrit{3} & $\forall\exists$ & 7 & ✓
& ❗ & 🔋 & \fa{9.0} & \na & \na & \na \\
\cdashline{2-11}[.4pt/.8pt]
& \rowstrut{}\ctrlspecone{2}{0} & $\exists\forall$ & 3 & ✗
& 24 & 0.7 & \fa{0.3} & \na & \na & \na \\
\cdashline{2-11}[.4pt/.8pt]
& \rowstrut{}\ctrlspecone{4}{0} & $\exists\forall$ & 7 & ✓
& ⏱ & ⏱ & \fa{0.4} & \na & \na & \fg{0.7} \\
\cdashline{2-11}[.4pt/.8pt]
& \rowstrut{}\ctrlspecone{4}{2} & $\exists\forall$ & 11 & ✓
& 🔋 & 🔋 & \fa{14} & \na & \na & \fg{86} \\
\cdashline{2-11}[.4pt/.8pt]
& \rowstrut{}\ctrlspectwo{1}{1} & $\exists\forall$ & 3 & ✗
& 89 & 3.4 & \fa{0.3} & \na & \na & \na \\
\cdashline{2-11}[.4pt/.8pt]
& \rowstrut{}\ctrlspectwo{1}{2} & $\exists\forall$ & 4 & ✓
& ⏱ & 1.0 & \fa{0.3} & \na & \na & \fa{0.3} \\
\cdashline{2-11}[.4pt/.8pt]
& \rowstrut{}\ctrlspectwo{4}{4} & $\exists\forall$ & 16 & ✓
& 🔋 & 🔋 & 🔋 & \na & \na & \fa{2.1} \\
\cdashline{2-11}[.4pt/.8pt]
& \rowstrut{}\robotplan{3}{1} & $\exists\forall$ & 3 & ✓
& 4.1 & 2.8 & \fg{0.4} & \na & \na & \fa{0.2} \\
\cdashline{2-11}[.4pt/.8pt]
& \rowstrut{}\robotplan{3}{2} & $\exists\forall^{2}$ & 5 & ✓
& 🔋 & 5.4 & \fg{0.4} & \na & \na & \fa{0.3} \\
\cdashline{2-11}[.4pt/.8pt]
& \rowstrut{}\robotplan{3}{3} & $\exists\forall^{3}$ & 6 & ✗
& 🔋 & 215 & \fa{0.4} & \na & \na & \na \\
\cdashline{2-11}[.4pt/.8pt]
& \rowstrut{}\robotplan{4}{1} & $\exists\forall$ & 4 & ✓
& 122 & 🔋 & \fa{0.4} & \na & \na & \fg{0.5} \\
\cdashline{2-11}[.4pt/.8pt]
& \rowstrut{}\robotplan{4}{2} & $\exists\forall^{2}$ & 5 & ✓
& 🔋 & ⏱ & \fg{0.4} & \na & \na & \fa{0.3} \\
\cdashline{2-11}[.4pt/.8pt]
& \rowstrut{}\robotplan{4}{3} & $\exists\forall^{3}$ & 7 & ✓
& ⏱ & ⏱ & \fa{0.5} & \na & \na & \fa{0.5} \\
\cdashline{2-11}[.4pt/.8pt]
& \rowstrut{}\robotplan{4}{4} & $\exists\forall^{4}$ & 9 & ✗
& ⏱ & ⏱ & \fa{0.5} & \na & \na & \na \\
\bottomrule
\end{tabular}
\end{table}

\section{Conclusion}\label{sec:conclusion}

\tool{} keeps \autohyper's algorithm and changes what each of its steps is handed.
The models keep their frozen and free variables symbolic on the edges of a quotiented explicit system.
The formula is reduced to fewer APs and decomposed into parts, whose products on the parts are quotiented as they are built.
The final inclusion check hands \kofola{} its left operand as parts, and every witness is lifted back to the original models.

The benchmark suites evaluated here are dominated by formulas with at most one alternation, so complementation is rarely exercised, since for those cases \kofola{} runs only as an inclusion checker.
Natural future work would be to evaluate \tool{} against a larger suite of multi-alternation benchmarks.

\subsubsection*{Data-Availability Statement.}
The latest version of \tool{} and all its supporting material can be found at the repository \url{https://github.com/haslab/OmegaHyper}.
It contains the complete source code of \tool{}, all benchmarks from the three suites, and automated scripts that run every tool and replicates the results found in \Cref{tab:results}. A self-contained docker image is also provided. \tool{} extends \autohyper{} and inherits its GPL~v3 license.

\bibliographystyle{splncs04}
\bibliography{refs}

@inproceedings{ClarksonFKMRS14,
  author    = {Michael R. Clarkson and Bernd Finkbeiner and Masoud Koleini and Kristopher K. Micinski
               and Markus N. Rabe and C{\'{e}}sar S{\'{a}}nchez},
  title     = {Temporal Logics for Hyperproperties},
  booktitle = {POST},
  series    = {LNCS},
  volume    = {8414},
  pages     = {265--284},
  publisher = {Springer},
  year      = {2014},
  doi       = {10.1007/978-3-642-54792-8_15}
}

@article{symmetry,
  author    = {Alice Miller and Alastair Donaldson and Muffy Calder},
  title     = {Symmetry in Temporal Logic Model Checking},
  journal   = {{ACM} Comput. Surv.},
  volume    = {38},
  number    = {3},
  pages     = {8},
  year      = {2006},
  doi       = {10.1145/1132960.1132962}
}

@inproceedings{FinkbeinerRS15,
  author    = {Bernd Finkbeiner and Markus N. Rabe and C{\'{e}}sar S{\'{a}}nchez},
  title     = {Algorithms for Model Checking {HyperLTL} and {HyperCTL*}},
  booktitle = {CAV},
  series    = {LNCS},
  volume    = {9206},
  pages     = {30--48},
  publisher = {Springer},
  year      = {2015},
  doi       = {10.1007/978-3-319-21690-4_3}
}

@inproceedings{CoenenFST19,
  author    = {Norine Coenen and Bernd Finkbeiner and C{\'{e}}sar S{\'{a}}nchez and Leander Tentrup},
  title     = {Verifying Hyperliveness},
  booktitle = {CAV},
  series    = {LNCS},
  volume    = {11561},
  pages     = {121--139},
  publisher = {Springer},
  year      = {2019},
  doi       = {10.1007/978-3-030-25540-4_7}
}

@article{beutner2025autohyper,
  author    = {Raven Beutner and Bernd Finkbeiner},
  title     = {{AutoHyper}: Leveraging Language Inclusion Checking for Hyperproperty Model-Checking},
  journal   = {Int. J. Softw. Tools Technol. Transf.},
  year      = {2025},
  note      = {Online first},
  doi       = {10.1007/s10009-025-00801-5}
}

@inproceedings{BeutnerF23,
  author    = {Raven Beutner and Bernd Finkbeiner},
  title     = {{AutoHyper}: Explicit-State Model Checking for {HyperLTL}},
  booktitle = {TACAS},
  series    = {LNCS},
  volume    = {13993},
  pages     = {145--163},
  publisher = {Springer},
  year      = {2023},
  doi       = {10.1007/978-3-031-30823-9_8}
}

@inproceedings{HsuSB21,
  author    = {Tzu{-}Han Hsu and C{\'{e}}sar S{\'{a}}nchez and Borzoo Bonakdarpour},
  title     = {Bounded Model Checking for Hyperproperties},
  booktitle = {TACAS},
  series    = {LNCS},
  volume    = {12651},
  pages     = {94--112},
  publisher = {Springer},
  year      = {2021},
  doi       = {10.1007/978-3-030-72016-2_6}
}

@inproceedings{hyperqb2,
  author    = {Tzu{-}Han Hsu and Milad Rabizadeh and Kenneth Rogale and Fedor Filippov
               and Marco A. de Oliveira Batista and Borzoo Bonakdarpour},
  title     = {{HyperQB} 2.0: A Bounded Model Checker for Hyperproperties},
  booktitle = {CAV},
  series    = {LNCS},
  volume    = {16682},
  pages     = {511--525},
  publisher = {Springer},
  year      = {2026},
  doi       = {10.1007/978-3-032-32519-8_26}
}

@inproceedings{HsuSSB23,
  author    = {Tzu{-}Han Hsu and C{\'{e}}sar S{\'{a}}nchez and Sarai Sheinvald and Borzoo Bonakdarpour},
  title     = {Efficient Loop Conditions for Bounded Model Checking Hyperproperties},
  booktitle = {TACAS},
  series    = {LNCS},
  volume    = {13993},
  pages     = {66--84},
  publisher = {Springer},
  year      = {2023},
  doi       = {10.1007/978-3-031-30823-9_4}
}

@inproceedings{hyperlasso,
  author    = {Alcino Cunha and Hugo Pacheco and Nuno Macedo},
  title     = {{HyperLasso}: Bounded Model Checking of $\forall^+\exists^+$-Liveness Hyperproperties},
  booktitle = {CAV},
  series    = {LNCS},
  volume    = {16682},
  pages     = {473--496},
  publisher = {Springer},
  year      = {2026},
  doi       = {10.1007/978-3-032-32519-8_24}
}

@article{hypersmv,
  author    = {Nuno Macedo and Hugo Pacheco},
  title     = {Model Checking of Hyperproperties for High-Level Relational Models},
  journal   = {{ACM} Trans. Softw. Eng. Methodol.},
  year      = {2026},
  doi       = {10.1145/3845615}
}

@inproceedings{Duret-LutzRCRAS22,
  author    = {Alexandre Duret{-}Lutz and Etienne Renault and Maximilien Colange and Florian Renkin
               and Alexandre Gbaguidi Aisse and Philipp Schlehuber{-}Caissier and Thomas Medioni
               and Antoine Martin and J{\'{e}}r{\^{o}}me Dubois and Cl{\'{e}}ment Gillard
               and Henrich Lauko},
  title     = {From {Spot} 2.0 to {Spot} 2.10: What's New?},
  booktitle = {CAV},
  series    = {LNCS},
  volume    = {13372},
  pages     = {174--187},
  publisher = {Springer},
  year      = {2022},
  doi       = {10.1007/978-3-031-13188-2_9}
}

@inproceedings{HavlenaLLST23,
  author    = {Vojt{\v{e}}ch Havlena and Ond{\v{r}}ej Leng{\'{a}}l and Yong Li and Barbora
               {\v{S}}mahl{\'{i}}kov{\'{a}} and Andrea Turrini},
  title     = {Modular Mix-and-Match Complementation of {B}{\"{u}}chi Automata},
  booktitle = {TACAS},
  series    = {LNCS},
  volume    = {13993},
  pages     = {249--270},
  publisher = {Springer},
  year      = {2023},
  doi       = {10.1007/978-3-031-30823-9_13}
}

@inproceedings{DoveriGM22,
  author    = {Kyveli Doveri and Pierre Ganty and Nicolas Mazzocchi},
  title     = {{FORQ}-Based Language Inclusion Formal Testing},
  booktitle = {CAV},
  series    = {LNCS},
  volume    = {13372},
  pages     = {109--129},
  publisher = {Springer},
  year      = {2022},
  doi       = {10.1007/978-3-031-13188-2_6}
}

@article{ClarksonS10,
  author  = {Michael R. Clarkson and Fred B. Schneider},
  title   = {Hyperproperties},
  journal = {J. Comput. Secur.},
  volume  = {18},
  number  = {6},
  pages   = {1157--1210},
  year    = {2010},
  doi       = {10.3233/JCS-2009-0393}
}

@article{Finkbeiner23,
  author  = {Bernd Finkbeiner},
  title   = {Logics and Algorithms for Hyperproperties},
  journal = {{ACM} {SIGLOG} News},
  volume  = {10},
  number  = {2},
  pages   = {4--23},
  year    = {2023},
  doi       = {10.1145/3610392.3610394}
}

@inproceedings{BeutnerF22csf,
  author    = {Raven Beutner and Bernd Finkbeiner},
  title     = {Prophecy Variables for Hyperproperty Verification},
  booktitle = {CSF},
  pages     = {471--485},
  publisher = {{IEEE}},
  year      = {2022},
  doi       = {10.1109/CSF54842.2022.9919658}
}

@inproceedings{BeutnerF22cav,
  author    = {Raven Beutner and Bernd Finkbeiner},
  title     = {Software Verification of Hyperproperties Beyond k-Safety},
  booktitle = {CAV},
  series    = {LNCS},
  volume    = {13371},
  pages     = {341--362},
  publisher = {Springer},
  year      = {2022},
  doi       = {10.1007/978-3-031-13185-1_17}
}

@inproceedings{BeutnerF23lpar,
  author    = {Raven Beutner and Bernd Finkbeiner},
  title     = {Model Checking {$\omega$}-Regular Hyperproperties with {AutoHyperQ}},
  booktitle = {LPAR},
  series    = {EPiC Series in Computing},
  volume    = {94},
  pages     = {23--35},
  publisher = {EasyChair},
  year      = {2023},
  doi       = {10.29007/1xjt}
}

@inproceedings{Beutner24,
  author    = {Raven Beutner},
  title     = {Automated Software Verification of Hyperliveness},
  booktitle = {TACAS},
  series    = {LNCS},
  volume    = {14571},
  pages     = {196--216},
  publisher = {Springer},
  year      = {2024},
  doi       = {10.1007/978-3-031-57249-4_10}
}

@article{DardinierLM24,
  author  = {Thibault Dardinier and Anqi Li and Peter M{\"{u}}ller},
  title   = {Hypra: A Deductive Program Verifier for Hyper {H}oare Logic},
  journal = {Proc. {ACM} Program. Lang.},
  volume  = {8},
  number  = {{OOPSLA2}},
  pages   = {1279--1308},
  year    = {2024},
  doi       = {10.1145/3689756}
}

@inproceedings{ItzhakySV24,
  author    = {Shachar Itzhaky and Sharon Shoham and Yakir Vizel},
  title     = {Hyperproperty Verification as {CHC} Satisfiability},
  booktitle = {ESOP},
  series    = {LNCS},
  volume    = {14577},
  pages     = {212--241},
  publisher = {Springer},
  year      = {2024},
  doi       = {10.1007/978-3-031-57267-8_9}
}

@article{CorrensonNFW24,
  author  = {Arthur Correnson and Tobias Nie{\ss}en and Bernd Finkbeiner and Georg Weissenbacher},
  title   = {Finding $\forall\exists$ Hyperbugs using Symbolic Execution},
  journal = {Proc. {ACM} Program. Lang.},
  volume  = {8},
  number  = {{OOPSLA2}},
  pages   = {1420--1445},
  year    = {2024},
  doi       = {10.1145/3689761}
}

@article{sebastiani2011hybrid,
  author    = {Roberto Sebastiani and Stefano Tonetta and Moshe Y. Vardi},
  title     = {Symbolic Systems, Explicit Properties: On Hybrid Approaches for {LTL} Symbolic Model Checking},
  journal   = {Int. J. Softw. Tools Technol. Transf.},
  volume    = {13},
  number    = {4},
  pages     = {319--335},
  publisher = {Springer},
  year      = {2011},
  doi = {10.1007/s10009-010-0168-4}
}

@inproceedings{HsuBFS23,
  author    = {Tzu{-}Han Hsu and Borzoo Bonakdarpour and Bernd Finkbeiner and C{\'{e}}sar S{\'{a}}nchez},
  title     = {Bounded Model Checking for Asynchronous Hyperproperties},
  booktitle = {TACAS},
  series    = {LNCS},
  volume    = {13993},
  pages     = {29--46},
  publisher = {Springer},
  year      = {2023},
  doi       = {10.1007/978-3-031-30823-9_2}
}

@inproceedings{BaumeisterCBFS21,
  author    = {Jan Baumeister and Norine Coenen and Borzoo Bonakdarpour and Bernd Finkbeiner and C{\'{e}}sar S{\'{a}}nchez},
  title     = {A Temporal Logic for Asynchronous Hyperproperties},
  booktitle = {CAV},
  series    = {LNCS},
  volume    = {12759},
  pages     = {694--717},
  publisher = {Springer},
  year      = {2021},
  doi       = {10.1007/978-3-030-81685-8_33}
}

@article{BeutnerF25fmsd,
  author  = {Raven Beutner and Bernd Finkbeiner},
  title   = {Predicate Abstraction for Hyperliveness Verification},
  journal = {Formal Methods Syst. Des.},
  volume  = {66},
  number  = {2},
  pages   = {238--277},
  year    = {2025},
  doi       = {10.1007/s10703-025-00482-5}
}

@article{BeutnerF25oopsla,
  author  = {Raven Beutner and Bernd Finkbeiner},
  title   = {Verifying Asynchronous Hyperproperties in Reactive Systems},
  journal = {Proc. {ACM} Program. Lang.},
  volume  = {9},
  number  = {{OOPSLA2}},
  pages   = {2227--2253},
  year    = {2025},
  doi       = {10.1145/3763130}
}

@article{KanellakisS90,
  author  = {Paris C. Kanellakis and Scott A. Smolka},
  title   = {{CCS} Expressions, Finite State Processes, and Three Problems of Equivalence},
  journal = {Inf. Comput.},
  volume  = {86},
  number  = {1},
  pages   = {43--68},
  year    = {1990},
  doi     = {10.1016/0890-5401(90)90025-D}
}

@book{ClarkeGP99,
  author    = {Edmund M. Clarke and Orna Grumberg and Doron A. Peled},
  title     = {Model Checking},
  publisher = {{MIT} Press},
  year      = {1999},
  isbn      = {978-0-262-03270-4}
}

@inproceedings{AlexajHHLLM26,
  author    = {Ondrej Alexaj and Vojt{\v{e}}ch Havlena and Luk{\'{a}}{\v{s}} Hol{\'{i}}k
               and Ond{\v{r}}ej Leng{\'{a}}l and Yong Li and Nicolas Mazzocchi},
  title     = {{Kofola} 1.0: A Modular Approach to $\omega$-Regular Complementation and Inclusion Checking},
  booktitle = {CAV},
  series    = {LNCS},
  volume    = {16682},
  pages     = {181--197},
  publisher = {Springer},
  year      = {2026},
  doi       = {10.1007/978-3-032-32519-8_10}
}

@inproceedings{AbdullaCCHHMV11,
  author    = {Parosh Aziz Abdulla and Yu{-}Fang Chen and Lorenzo Clemente and Luk{\'{a}}{\v{s}} Hol{\'{i}}k
               and Chih{-}Duo Hong and Richard Mayr and Tom{\'{a}}{\v{s}} Vojnar},
  title     = {Advanced {Ramsey}-Based {B}{\"{u}}chi Automata Inclusion Testing},
  booktitle = {CONCUR},
  series    = {LNCS},
  volume    = {6901},
  pages     = {187--202},
  publisher = {Springer},
  year      = {2011},
  doi       = {10.1007/978-3-642-23217-6_13}
}

@inproceedings{DoveriGPR21,
  author    = {Kyveli Doveri and Pierre Ganty and Francesco Parolini and Francesco Ranzato},
  title     = {Inclusion Testing of {B}{\"{u}}chi Automata Based on Well-Quasiorders},
  booktitle = {CONCUR},
  series    = {LIPIcs},
  volume    = {203},
  pages     = {3:1--3:22},
  publisher = {Schloss Dagstuhl -- Leibniz-Zentrum f{\"{u}}r Informatik},
  year      = {2021},
  doi       = {10.4230/LIPIcs.CONCUR.2021.3}
}

@inproceedings{LiSTCX19,
  author    = {Yong Li and Xuechao Sun and Andrea Turrini and Yu{-}Fang Chen and Junnan Xu},
  title     = {{ROLL} 1.0: $\omega$-Regular Language Learning Library},
  booktitle = {TACAS},
  series    = {LNCS},
  volume    = {11427},
  pages     = {365--371},
  publisher = {Springer},
  year      = {2019},
  doi       = {10.1007/978-3-030-17462-0_23}
}

@article{BauchHB16,
  author  = {Petr Bauch and Vojt{\v{e}}ch Havel and Ji{\v{r}}{\'{\i}} Barnat},
  title   = {Control Explicit-Data Symbolic Model Checking},
  journal = {{ACM} Trans. Softw. Eng. Methodol.},
  volume  = {25},
  number  = {2},
  pages   = {1--48},
  year    = {2016},
  doi     = {10.1145/2888393}
}

@inproceedings{SmithV98,
  author    = {Geoffrey Smith and Dennis M. Volpano},
  title     = {Secure Information Flow in a Multi-threaded Imperative Language},
  booktitle = {POPL},
  pages     = {355--364},
  publisher = {ACM},
  year      = {1998},
  doi       = {10.1145/268946.268975}
}

\end{document}